\documentclass[subeqn,12pt, a4paper]{article} 
\usepackage{amssymb,amsmath,amsfonts,amsthm, amscd, mathrsfs,helvet}
\usepackage{bm, stmaryrd}
\usepackage{graphicx,verbatim}
\usepackage[usenames, dvipsnames]{color}
\usepackage{psfrag}
\usepackage[all]{xy}
\usepackage{tikz}
\usetikzlibrary{arrows,matrix,decorations.markings,shapes.geometric}
\usepackage[normalem]{ulem}
\usepackage{xspace}
\usepackage{srcltx}
\usepackage{pdfpages}
\usepackage[nosort]{cite}

\theoremstyle{plain}
\newtheorem{theorem}{Theorem}[section]
\newtheorem*{theorem*}{Theorem}

\newtheorem*{proposition*}{Proposition}

\newtheorem{lemma}[theorem]{Lemma}

\newcommand{\tensor}[1]{{\bf \underline{#1}}}

\definecolor{brightBlue}{rgb}{0,0,1}

\definecolor{Violet}{rgb}{0.47,0,1}

\DeclareMathOperator{\res}{res}
 \DeclareMathOperator{\End}{End}
\DeclareMathOperator{\ad}{ad}
\DeclareMathOperator{\Ad}{Ad}

\def\b{\mathfrak{b}}

\def\f{\mathfrak{f}}
\def\g{\mathfrak{g}}
\def\h{\mathfrak{h}}

\def\n{\mathfrak{n}}
\def\p{\partial}

\def\s{\sigma}

\def\gd{\mathfrak{g}_{\rm diag}}

\newcommand{\Id}{\text{Id}}
\newcommand{\U}{\mathscr{U}}
\newcommand{\Lc}{\mathcal{L}}
\newcommand{\Rc}{\mathcal{R}}
\newcommand{\Pc}{\mathcal{P}}
\newcommand{\Qc}{\mathcal{Q}}

\newcommand{\Diff}[1]{\text{Diff}(#1)}
\newcommand{\Vect}[1]{\text{Vect}(#1)}
\newcommand{\Pexp}{\text{P}\overleftarrow{\text{exp}}}
\newcommand{\ti}[1]{_{\bm{\underline{#1}}}}

\newcommand{\longhookrightarrow}{\lhook\joinrel\relbar\joinrel\rightarrow}

\def\C{\mathbb{C}}

\def\L{\mathcal{L}}

\def\N{\mathbb{N}}

\def\1{\tensor{1}}
\def\2{\tensor{2}}
\def\3{\tensor{3}}
\def\4{\tensor{4}}

\def\beq{\begin{equation}}
\def\eeq{\end{equation}}
\def\beqz{\begin{equation*}}
\def\eeqz{\end{equation*}}
\def\bea{\begin{eqnarray}}
\def\eea{\end{eqnarray}}
\def\dd{d}

\newcommand{\pcm}{principal chiral model\xspace}

\numberwithin{equation}{section}

\begin{document}

\begin{center}
\vspace*{2em}
{\large\bf  On $q$-deformed symmetries as Poisson-Lie symmetries \\[1mm]
and application to Yang-Baxter type models}\\
\vspace{1.5em}
F. Delduc$\,{}^1$, S. Lacroix$\,{}^1$, M. Magro$\,{}^1$, B. Vicedo$\,{}^2$

\vspace{1em}
\begingroup\itshape
{\it 1) 
Univ Lyon, Ens de Lyon, Univ Claude Bernard, CNRS, Laboratoire de Physique, \\
F-69342 Lyon, France} \\
\vspace{1em}
{\it 2) School of Physics, Astronomy and Mathematics,
University of Hertfordshire,}\\
{\it College Lane,
Hatfield AL10 9AB,
United Kingdom}
\par\endgroup
\vspace{1em}
\begingroup\ttfamily
Francois.Delduc@ens-lyon.fr, Sylvain.Lacroix@ens-lyon.fr, Marc.Magro@ens-lyon.fr, Benoit.Vicedo@gmail.com
\par\endgroup
\vspace{1.5em}
\end{center}

\begin{abstract}
Yang-Baxter type models are integrable deformations of integrable field theories, such as the principal chiral model on a Lie group $G$ or $\sigma$-models on (semi-)symmetric spaces $G/F$. The deformation has the effect of breaking the global $G$-symmetry of the original model, replacing the associated set of conserved charges by ones whose Poisson brackets are those of the $q$-deformed Poisson-Hopf algebra $\U_q(\g)$. Working at the Hamiltonian level, we show how this $q$-deformed Poisson algebra originates from a Poisson-Lie $G$-symmetry. The theory of Poisson-Lie groups and their actions on Poisson manifolds, in particular the formalism of the non-abelian moment map, is reviewed. For a coboundary Poisson-Lie group $G$, this non-abelian moment map must obey the Semenov-Tian-Shansky bracket on the dual group $G^*$, up to terms involving central quantities. When the latter vanish, we develop a general procedure linking this Poisson bracket to the defining relations of the Poisson-Hopf algebra $\U_q(\g)$, including the $q$-Poisson-Serre relations. We consider reality conditions leading to $q$ being either real or a phase. We determine the non-abelian moment map for Yang-Baxter type models. This enables to compute the corresponding action of $G$ on the fields parametrising the phase space of these models.
\end{abstract}

\section{Introduction}

Integrable field theories of so-called Yang-Baxter type arise as integrable deformations of different well known integrable $\sigma$-models. The name originates from the appearance of an $R$-matrix, a solution of the so-called modified classical Yang-Baxter equation (mCYBE) on a Lie algebra $\g$ which reads
\begin{equation*}
[RX, RY] - R \big( [RX, Y] + [X, RY] \big) = - c^2 [X, Y],
\end{equation*}
for any $X, Y \in \g$ and with $c \neq 0$, in the action of all these models. Without loss of generality we can assume that $c = 1$ or $c = i$, referred to as the split and non-split cases respectively. The first example 
of such a model, coined the Yang-Baxter $\sigma$-model, was constructed in 
\cite{Klimcik:2002zj,Klimcik:2008eq} by C. {Klim$\check{\text{c}}$\'{\i}k} as a one-parameter 
deformation of the principal chiral model. This was later generalised to deformations of 
symmetric \cite{Delduc:2013fga} as well as semi-symmetric space $\sigma$-models 
\cite{Delduc:2013qra,Delduc:2014kha}. Other $\sigma$-models of Yang-Baxter type include the 
bi-Yang-Baxter $\sigma$-model which gives a two-parameter deformation of the principal 
chiral model \cite{Klimcik:2008eq,Klimcik:2014bta}. The classical integrability of these 
various Yang-Baxter type models at the Hamiltonian level was proved in 
\cite{Delduc:2013fga,Delduc:2014kha,Delduc:2015xdm}. We note that the construction of 
Yang-Baxter type models can equally be applied to solutions of the classical Yang-Baxter 
equation (the case $c = 0$) \cite{Kawaguchi:2014qwa}, resulting in a different class of 
models \cite{Kawaguchi:2014fca,Matsumoto:2015jja,vanTongeren:2015soa} which may be referred to 
as ``homogeneous'' Yang-Baxter type models. However, these models have very different properties (see \emph{e.g.} \cite{Vicedo:2015pna}) to those obtained from solutions of the mCYBE and so we will not consider such models here.  

\medskip

A prominent feature shared by all Yang-Baxter type models built from solutions of the mCYBE is that they are characterised by a $q$-deformation of the global symmetry algebra of the original ``undeformed'' integrable $\sigma$-model. To illustrate this point let us take the \pcm on a real semisimple Lie group $G$ with Lie algebra $\g = \textup{Lie}\; G$ as an example. This model is invariant under a global $G \times G$ symmetry acting by left and right multiplications on its group valued field $g(\sigma, \tau)$. To describe this statement in the Hamiltonian formalism, let $M$ denote the phase space of the model with Poisson bracket $\{ \cdot, \cdot \}$. The left and right actions of $G$ on $M$ each admit a moment map $Q : M \to \g^\ast$ such that:
\begin{itemize}
  \item[$(1)$] the infinitesimal symmetry with parameter $\epsilon \in \g$ reads $\delta_\epsilon f = \langle \epsilon, \{ Q, f \} \rangle$ for any function $f : M \to \mathbb{R}$ on phase space, where $\langle \cdot, \cdot \rangle$ denotes the pairing between $\g$ and $\g^\ast$, and
  \item[$(2)$] writing $Q = Q^a T_a$ where $T^a$ and $T_a$ are dual bases of $\g$ and $\g^\ast$, we have $\{ Q^a, Q^b \} = f^{ab}_{\;\;\; c} \, Q^c$ with $f^{ab}_{\;\;\; c}$ denoting the structure constants of $\g$ with respect to $T^a$, \emph{i.e.} $[T^a, T^b] = f^{ab}_{\;\;\; c} \,T^c$.
\end{itemize}
Moreover, both the left and right actions of $G$ on $M$ are symmetries of the principal chiral model since they leave the Hamiltonian $H$ invariant, \emph{i.e.} $\delta_\epsilon H = 0$ for all $\epsilon \in \g$. For the Yang-Baxter $\sigma$-model it was shown in \cite{Delduc:2013fga} that the Poisson brackets from $(2)$ satisfied by the charges of the right $G$-symmetry get deformed to those of a Poisson-Hopf algebra $\U_q(\g)$ \cite{Ballesteros_2009,Kawaguchi:2012ve}, where $q$ is a function of the deformation parameter. The latter can be defined as a semiclassical limit of the quantum group $U_{\widehat q}(\g)$ with $\widehat q = q^\hbar$. This fact was first shown for the Yang-Baxter $\sigma$-model with $G = SU(2)$ in \cite{Kawaguchi:2011pf,Kawaguchi:2012gp}. Likewise, in the case of the bi-Yang-Baxter $\sigma$-model, the Poisson algebra of the global left and right $G$-symmetries get deformed instead to $\U_{q_l}(\g)$ and $\U_{q_r}(\g)$ algebras, where $q_l$ and $q_r$ are functions of the two deformation parameters. The latter were conjectured in \cite{Hoare:2014oua} and proved in \cite{Delduc:2015xdm}.

\medskip

One of the motivations for the present work is to address the following question: what is the infinitesimal transformation associated with the $q$-deformed symmetry of Yang-Baxter type models? In other words, what is the analog of the property $(1)$ above after deforming? Before answering this question it is worth recalling first that the phase space of a Yang-Baxter type deformation is the same as the phase space of the original ``undeformed'' integrable model \cite{Vicedo:2015pna}, which we shall keep calling $M$ as in the above example.
The correct framework for describing the symmetries of Yang-Baxter type models is that of Poisson-Lie groups and Poisson-Lie symmetries. Indeed, the effect of the deformation will be to promote the Lie group $G$, whose action on $M$ is Hamiltonian in the sense of property $(1)$ above, to the status of a Poisson-Lie group, whose action on $M$ is also Hamiltonian but in the Poisson-Lie group sense.

\medskip

Recall that a Poisson-Lie group\footnote{There are many references on Poisson-Lie groups. For the aspects reviewed in the present article, we mainly refer to the articles \cite{Drinfeld:1986in,Drinfeld:1983ky,SemenovTianShansky:1985my,Semenov-nonv,luweinstein1990_a,lu_1990_phd,Babelon:1991ah,Falceto:1992bf,Feher:2002fx} and to the books \cite{Babebook,Chari_Pressley_1994}. Further references may be found in \cite{Kosmann-Schwarzbach2004}.} $G$ is a Lie group equipped with a Poisson bracket $\{\cdot,\cdot\}_G$ which is compatible with the multiplication. Its linearisation at the identity endows the dual $\mathfrak{g}^\ast$ of its Lie algebra $\g$ with a natural Lie bracket $[ \cdot,\cdot]_\ast$. The dual group $G^\ast$ is defined as the corresponding connected and simply connected Lie group with $\textup{Lie}\; G^\ast = \g^\ast$. The action of the Poisson-Lie group $(G, \{\cdot,\cdot\}_G)$ on $M$ is called Hamiltonian if there exists a non-abelian moment map $\Gamma : M \to G^\ast$ such that the infinitesimal variation of any function $f : M \to \mathbb{R}$ is given by 
\beq \label{def-delta-intro}
\delta_\epsilon f = -\left\langle \epsilon, \Gamma^{-1}\left\lbrace \Gamma, f \right\rbrace \right\rangle
\eeq
where the parameter $\epsilon$ takes value in $\g$. Note that if the Poisson bracket $\{\cdot,\cdot\}_G$ is trivial then so is the Lie bracket $[ \cdot,\cdot]_\ast$, in which case the group $G^\ast$ becomes abelian. As a consequence, writing $\Gamma = \exp(-Q)$ with $Q : M \to \g^\ast$, one recovers the standard infinitesimal symmetry generated by the moment map $Q$ as in $(1)$ above. We therefore expect the Poisson bracket $\{ \cdot, \cdot \}_G$ to be proportional to the deformation parameter.

\medskip

If the deformation is to promote $G$ to a Poisson-Lie group, the question is with which Poisson structure $\{ \cdot, \cdot \}_G$? In the context of Yang-Baxter type deformations, which are associated with skew-symmetric $R$-matrices, one can define a natural coboundary Poisson-Lie group structure on $G$ given by the Sklyanin bracket
\begin{equation*}
\left\{ x\ti{1}, x\ti{2} \right\}_G = \gamma \left[R\ti{12}, x\ti{1}x\ti{2} \right],
\end{equation*}
where $x \in G$ and $\gamma$ plays the role of the real deformation parameter. In particular, we note that the Poisson bracket becomes trivial in the undeformed limit $\gamma = 0$. In the present case, the Lie algebra $(\g^\ast, [\cdot, \cdot]_\ast)$ defined above is naturally isomorphic to the Lie algebra $(\g, [\cdot, \cdot]_R)$, which we denote $\g_R$, where
\begin{equation*}
[X,Y]_R = \gamma \big([RX,Y]+[X,RY]\big),
\end{equation*}
for all $X, Y \in \g$. Note that this defines a Lie bracket by virtue of the mCYBE. Correspondingly, the dual group $G^\ast$ is isomorphic to the connected and simply connected Lie group $G_R$ with Lie algebra $\g_R$. Alternatively, depending on whether the $R$-matrix is a split or non-split solution of the mCYBE, the Lie algebra $(\g^\ast, [\cdot, \cdot]_\ast)$ may equally be realised as a Lie subalgebra of the real double $\g \oplus \g$ which we denote $\g_{DR}$, or as Lie subalgebras $\g_\pm$ of the complexification $\g^\C$, respectively. Let $G_{DR}$ be the corresponding Lie subgroup of $G \times G$ and $G_\pm$ the Lie subgroups of the complexification $G^\C$. Both $G_{DR}$ and $G_\pm$ become Poisson-Lie groups themselves when equipped with the Semenov-Tian-Shansky Poisson bracket
\begin{equation*}
\left\{ x^{\pm}\ti{1}, x^{\pm}\ti{2} \right\}_{G^\ast} = \gamma \left[R\ti{12}, x^{\pm}\ti{1}x^{\pm}\ti{2} \right], \qquad
\left\{ x^{\pm}\ti{1}, x^{\mp}\ti{2} \right\}_{G^\ast} = \gamma \left[R^{\pm}\ti{12}, x^{\pm}\ti{1}x^{\mp}\ti{2} \right],
\end{equation*}
where $(x^+, x^-) \in G_{DR}$ or $x^\pm \in G_\pm$ and we have defined $R^\pm = R \pm c \, \textup{Id}$.

\medskip

We are now in a position to state what the generalisation of the above two properties $(1)$ and $(2)$ should be for Yang-Baxter type models. Namely, the action of the coboundary Poisson-Lie group $(G, \{ \cdot, \cdot \}_G)$ on $M$ should admit a non-abelian moment map, which can be described as a map $(\Gamma^+, \Gamma^-) : M \to G_{DR}$ or $\Gamma^\pm : M \to G_\pm$, such that:
\begin{itemize}
  \item[$(1')$] the infinitesimal symmetry with parameter $\epsilon \in \g$ reads
\begin{equation} \label{def2-delta-intro}
\delta_\epsilon f = -\frac{1}{2 c \gamma} \kappa \Bigl( \epsilon, (\Gamma^+)^{-1}
\left\lbrace \Gamma^+, f \right\rbrace - (\Gamma^-)^{-1}\left\lbrace \Gamma^-, f \right\rbrace \Bigr)
\end{equation}
for any function $f : M \to \mathbb{R}$ on phase space, where $\kappa(\cdot,\cdot)$ is the Killing form on $\g$, and
  \item[$(2')$] the non-abelian moment map is a Poisson map, in the sense that
\begin{equation} \label{STS-intro}
\left\lbrace \Gamma^\pm\ti{1}, \Gamma^\pm\ti{2} \right\rbrace  =  
\gamma \left[ R\ti{12}, \Gamma^\pm\ti{1} \Gamma^\pm\ti{2} \right], \qquad
\left\lbrace \Gamma^\pm\ti{1}, \Gamma^\mp\ti{2} \right\rbrace = 
\gamma \left[ R^\pm\ti{12}, \Gamma^\pm\ti{1} \Gamma^\mp\ti{2} \right].
\end{equation}
\end{itemize}
Moreover, the action of the Poisson-Lie group $(G,\{\cdot,\cdot\}_G)$ on $M$ is a symmetry if it preserves the Hamiltonian, \emph{i.e.} $\delta_\epsilon H = 0$ for all $\epsilon \in \g$, or equivalently if the non-abelian moment map is conserved. For the standard choice of $R$-matrix, in both the split and non-split cases, the data of the non-abelian moment map $\Gamma^\pm$ is equivalently specified by a collection of charges $Q_i^H$ for $i = 1, \ldots, l = \text{rank}\, \g$ and $Q^E_\alpha$ for each root $\alpha \in \h^\ast$ of $\g$. We will show that the Poisson bracket relations \eqref{STS-intro} on the non-abelian moment map $\Gamma^\pm$ implies that these charges satisfy the defining Poisson bracket relations, including the $q$-Poisson-Serre relations, of the Poisson-Hopf algebra $\U_q(\g)$ where $q = e^{- i c \gamma}$.

\medskip

In order to justify our claim that the $q$-deformed symmetries of Yang-Baxter type models indeed correspond to Poisson-Lie group symmetries in the above sense, it therefore remains to identify the corresponding non-abelian moment map $\Gamma : M \to G^\ast$. Now Yang-Baxter type models can be obtained by deforming a double pole of the twist function of the undeformed integrable models into a pair of simple poles at $\lambda_\pm$ \cite{Vicedo:2015pna}. In this context, focusing on the case of a non-split $R$-matrix, we will show that the gauge transformed monodromy matrix evaluated at the pair of points $\lambda_\pm$, namely $\Gamma^\pm = T^g(\lambda_\pm)$, satisfy the Poisson bracket relations \eqref{STS-intro} and thus provide the desired non-abelian moment map.
 
\medskip

This achieves the main goal of this article, namely the generic identification of $q$-deformed symmetries analysed in \cite{Delduc:2013fga} with Poisson-Lie symmetries in Yang-Baxter type models. It is worth emphasising that the study of Poisson-Lie symmetries in the context of various $\sigma$-models of Yang-Baxter type has been carried out before in \cite{Klimcik:2002zj,Hoare:2014oua}.
In fact, the Yang-Baxter $\sigma$-model and the bi-Yang-Baxter $\sigma$-model were both originally defined 
in \cite{Klimcik:2002zj, Klimcik:2008eq} so as to possess, by their very construction, Poisson-Lie symmetries in the sense of \cite{Klimcik:1995ux,Klimcik:1996br}. By contrast, we wish to stress that the analysis carried out in the present article applies to all Yang-Baxter type models and, most importantly, operates at the Hamiltonian level. The latter is the right setting not only for discussing the Poisson brackets \eqref{STS-intro} of the non-abelian moment map but also to express the sought after infinitesimal transformation \eqref{def2-delta-intro} it generates. The variations of the canonical fields of the model are of the same form as the variations in the undeformed model but where the transformation parameters are suitably dressed by non-local expressions in the fields.
 
\medskip

The plan of this article is the following. In section \ref{sec: PL Dr double} we begin by reviewing the definitions of Lie bialgebras, Poisson-Lie groups and Drinfel'd doubles. In the next section we go on to recall what is meant by the action of a Poisson-Lie group on a Poisson manifold $M$. The action is called Hamiltonian when the corresponding infinitesimal transformation is generated by a non-abelian moment map $\Gamma : M \to G^\ast$ as in \eqref{def-delta-intro}. The requirement that this be a Lie algebra action is shown to imply that the lift
\begin{equation*}
M \longrightarrow G^\ast \longhookrightarrow DG
\end{equation*}
of $\Gamma$ to the Drinfel'd double $DG$ is a Poisson map if we equip $DG$ with the Sklyanin bracket associated with the canonical $\mathcal R$-matrix, up to terms involving central quantities.

In section \ref{Sec:Cobound} we turn to the study of coboundary Poisson-Lie groups, discussing in parallel the cases of split and non-split $R$-matrices. In this setting we recall the different realisations of the dual group $G^\ast$ which in later sections will provide us with useful ways of describing the non-abelian moment map $\Gamma$. We focus on the Poisson-Lie group structure on $G$ defined by the Sklyanin bracket. The dual space $\g^\ast$ is naturally a Lie algebra which is isomorphic in this case to $\g_R$, the vector space $\g$ equipped with the $R$-bracket. This provides a first concrete realisation of $G^\ast$ as the Lie group $G_R$ associated to $\g_R$. We denote the corresponding description of the non-abelian moment map by
\begin{equation*}
\Gamma_R : M \longrightarrow G_R.
\end{equation*}
Next, we recall that the Drinfel'd double $D\g$ of a coboundary Lie bialgebra defined by a split $R$-matrix is canonically isomorphic as a Lie algebra to the real double $\g \oplus \g$. Likewise, in the case of a non-split $R$-matrix the Drinfel'd double $D\g$ is isomorphic as a Lie algebra to the complexification $\g^\C$. The Lie algebra $\g^\ast \simeq \g_R$ may then be realised in the split case as a Lie subalgebra of $\g \oplus \g$ which is complementary to the diagonal subalgebra, or as a Lie subalgebra of $\g^\C$ complementary to the real subalgebra in the non-split case. This provides a second explicit realisation of the dual group $G^\ast$ as a subgroup of the Cartesian product $G \times G$ in the split case or of the complexification $G^\C$ in the non-split case. We denote the corresponding presentations of the non-abelian moment map respectively by
\begin{equation*}
(\Gamma^+, \Gamma^-) : M \longrightarrow G_{DR}, \qquad \text{and} \qquad
\Gamma^\pm : M \longrightarrow G_\pm.
\end{equation*}
We show that these are both Poisson maps if we equip $G_{DR}$ and $G_\pm$ with the Semenov-Tian-Shansky Poisson bracket, again with the possible addition of terms involving central quantities. In the remainder of the paper, however, we shall put these central quantities to zero.

Section \ref{Sec:qAlgebra} is devoted to showing that, when we choose the standard $R$-matrix in the split (resp. non-split) case, the Semenov-Tian-Shansky Poisson bracket on $G_{DR}$ (resp. $G_\pm$) corresponds precisely to the defining Poisson bracket relations of the Poisson-Hopf algebra $\U_q(\g)$ with $q = e^{- i \gamma}$ (resp. $q = e^{\gamma}$). Specifically, the choice of standard $R$-matrix means that $\Gamma^+$ and $\Gamma^-$ are elements of the positive and negative Borel subgroups of $G$ (resp. $G^\C$). To any choice of normal ordering on the set of positive roots of $\g$ we associate a parametrisation of $\Gamma^+$ and $\Gamma^-$ in terms of charges $Q^H_i$ for each Cartan direction and $Q^E_\alpha$ for each positive and negative root $\alpha \in \h^\ast$. We then show that imposing the Semenov-Tian-Shansky Poisson bracket on the pair $\Gamma^\pm$ implies that the charges $Q^H_i$ and $Q^E_\alpha$ satisfy all the relations of the Poisson algebra $\U_q(\g)$, including the $q$-Poisson-Serrre relations, as given in \cite{Delduc:2013fga}. We also discuss reality conditions suitable for the split and non-split cases.

Finally, the results from previous sections are applied in section \ref{sec: YB models} to discuss the $q$-deformed symmetries of Yang-Baxter type models. As explained above, we show that the non-abelian moment map in this case is given by the quantity $\Gamma^\pm = T^g(\lambda_\pm)$. We use this to compute the variations under the $q$-deformed symmetry of the canonical fields parametrising the phase space of these models. To end this section we also comment on the corresponding variation of the first order action of Yang-Baxter type models.

\section{Poisson-Lie groups and Drinfel'd doubles} \label{sec: PL Dr double}

In this section, we recall the main points of the general theory of Poisson-Lie groups and their link to Lie bialgebras, including the formulation in terms of Drinfel'd doubles.

\subsection{Poisson-Lie groups and Lie bialgebras}

A Poisson-Lie group is a real Lie group $G$ equipped with a Poisson bracket $\lbrace \cdot, \cdot \rbrace_G$ which is compatible with the multiplication $G\times G \to G$ in the sense that the latter is a Poisson map.

Consider the dual space $\g^*$ of the Lie algebra $\g$. As $\g \simeq T_eG$, any element in $\g^*$ can be realised as the differential $\dd_e f:T_eG \rightarrow \mathbb{R}$ of a smooth function $f : G \to \mathbb{R}$, taken at the identity $e$. Using this, we define a skew-symmetric product on $\g^*$ by
\begin{equation}\label{LieDual}
\left[ d_ef, d_eg \right]_* = d_e \left\lbrace f,g \right\rbrace_G.
\end{equation}
One can show that this product is well defined, \emph{i.e.} that the results only depend on $d_ef$ and $d_eg$ and not on the choice of $f$ and $g$. Using the Jacobi identity of the Poisson bracket, one finds that
\begin{equation*}
(\g^*,[\cdot,\cdot]_*)
\end{equation*}
is a Lie algebra. The Lie bracket $[\cdot,\cdot]_*$ can be seen as a skew-symmetric 
map $\delta^*:\g^* \otimes \g^* \rightarrow \g^*$. One can show that the dual map 
$\delta:\g \rightarrow \g \otimes \g$ is a 1-cocycle, \textit{i.e.} that $(\g,\g^*)$ is a Lie bialgebra (see for instance  \cite{Chari_Pressley_1994}). Conversely, from any Lie bialgebra, one can define a unique connected and simply connected Poisson-Lie group.

\subsection{Drinfel'd doubles}
\label{SubSec:Drinfeld}

Let $G$ be a Poisson-Lie group, with Lie bialgebra $(\g,\g^*)$. We define the Drinfel'd double of $\g$ as the vector space direct sum
\begin{equation*}
D\g = \g \oplus \g^*.
\end{equation*}
We will write $\iota$ and $\iota^*$ for the natural embeddings of $\g$ and $\g^*$ into $D\g$ and we will denote elements of $D\g$ as $(X,\lambda)$, where $X$ is in $\g$ and $\lambda$ is a linear form in $\g^*$. One can define a bilinear form on the double $D\g$ by
\begin{equation}\label{PairingDg}
\big\langle (X,\lambda) | (Y,\mu) \big\rangle = \langle X, \mu \rangle + \langle Y, \lambda \rangle = \mu(X) + \lambda(Y)
\end{equation}
for any $X, Y \in \g$ and $\lambda, \mu \in \g^\ast$, where $\langle \cdot, \cdot \rangle$ denotes the canonical pairing between $\g$ and $\g^*$. One then has the following result~\cite{Chari_Pressley_1994}:

\begin{theorem} \label{thm: Drin double}
There exists a unique Lie bracket $[\cdot,\cdot]_D$ on $D\g$ such that $\iota$ and $\iota^*$ are Lie homomorphisms from $\g$ and $\g^*$ to $D\g$, and such that the bilinear form $\langle \cdot | \cdot \rangle$ is $\ad$-invariant.
\end{theorem}

It is clear from the definition \eqref{PairingDg} that the subspaces $\iota(\g)$ and $\iota^\ast(\g^*)$ are both isotropic with respect to $\langle \cdot | \cdot \rangle$. The data $(D\g,\g,\g^*)$ therefore defines a Manin triple. Moreover, let us consider a basis $\lbrace T^a \rbrace$ of $\g$ and the dual basis $\lbrace T_a \rbrace$ of $\g^*$. Consider the element in $D\g\otimes D\g$ define by
\begin{equation}\label{RD}
\Rc^D\ti{12} = \sum_a \iota(T^a) \otimes \iota^*(T_a) - \iota^*(T_a) \otimes \iota(T^a),
\end{equation}
where we use the standard tensorial notation.
By the Adler-Kostant-Symes procedure, it is a skew-symmetric solution of the split modified classical Yang-Baxter equation on $D\g$, which in tensorial notation reads
\begin{equation*}
\left[ \Rc^D\ti{12},\Rc^D\ti{13} \right]_D + \left[ \Rc^D\ti{12},\Rc^D\ti{23} \right]_D + \left[ \Rc^D\ti{13},\Rc^D\ti{23} \right]_D = \left[ C^D\ti{12},C^D\ti{13} \right]_D,
\end{equation*}
with $C^D\ti{12}$ the quadratic Casimir in $D\g \otimes D\g$.

\section{Poisson-Lie actions}

In this section we study actions of Poisson-Lie groups on Poisson manifolds \textit{via} the non-abelian moment map formulation.

\subsection{Non-abelian moment map} \label{subsec: namm}

Let $G$ be a Poisson-Lie group and $M$ be a Poisson manifold, with Poisson bracket $\lbrace \cdot, \cdot \rbrace$. Let
\begin{equation*}
\rho : G \times M \longrightarrow M
\end{equation*}
be a smooth group action of $G$ on $M$. We say that $\rho$ is a Poisson-Lie action if it is a Poisson map, from $G \times M$, with the direct product Poisson structure, to $M$. The map $\rho$ can alternatively be seen as a group homomorphism from $G$ to $\Diff{M}$, the group of diffeomorphisms of $M$. Its differential at the identity induces a Lie algebra action
\begin{equation*}
\delta : \g \longrightarrow T_{\Id} \bigl(\Diff{M}\bigr) = \Vect{M},
\end{equation*}
on the space $\Vect{M}$ of vector fields on $M$. For $\epsilon\in\g$, the vector field $\delta_\epsilon$ acts naturally on any smooth function $f:M\rightarrow\mathbb{R}$. We consider the case where there exists a map
\begin{equation*}
\Gamma : M \longrightarrow G^*,
\end{equation*}
where the dual group $G^*$ is the connected and simply connected Lie group with Lie algebra $\g^*$, such that
\begin{equation}\label{MomentMap}
\delta_\epsilon f = -\left\langle \epsilon, \Gamma^{-1}\left\lbrace \Gamma, f \right\rbrace \right\rangle.
\end{equation}
The map $\Gamma$ is called the non-abelian moment map of the action of $G$ on $M$. If  $M$ is symplectic and simply connected, then such a map always exists. We can note here that $\Gamma$ is defined up to a left multiplication by a constant element in $G^*$. Conversely, every transformation of the form \eqref{MomentMap} preserves the Poisson bracket if the parameter $\epsilon$ has a non-zero bracket with itself, coming from the Poisson-Lie structure on $G$.

To illustrate this concept, let us investigate here the case of a usual Hamiltonian action of $G$ on $M$. For any fixed $g$, the action $\rho(g,\cdot)$ is then a canonical transformation on $M$. In other words, $\rho$ is a Poisson map for the trivial Poisson structure on $G$. The induced Lie bracket on $\g^*$ defined by \eqref{LieDual} is then trivial, so that the dual group $G^*$ is abelian. We write $\Gamma = \exp(-Q)$, with $Q: M \rightarrow \g^*$. As $G^*$ is abelian, the transformation \eqref{MomentMap} simply becomes
\begin{equation}\label{CanonicalCase}
\delta_\epsilon f = \left\langle \epsilon, \left\lbrace Q, f \right\rbrace \right\rangle.\vspace{-2pt}
\end{equation}
We recognize here the usual expression for a Hamiltonian action of $G$ on $M$, with $Q$ the moment map. When this action is a symmetry of a Hamiltonian system, decomposing $Q$ with respect to the dual basis of $\g^\ast$ as $Q=\sum_a Q^a T_a$, we obtain $\text{dim}\,G$ conserved charges $Q^a$.

\subsection{Poisson brackets of the non-abelian moment map}
\label{SubSec:PBGamma}

Let us recall that since $\rho$ is a Lie group action, $\delta$ is a Lie algebra action. In other words, $\delta$ is a Lie homomorphism which is to say that
\begin{equation}\label{ActionLie}
\left[ \delta_\epsilon,\delta_{\epsilon'} \right] = \delta_{[\epsilon,\epsilon']},
\end{equation}
for any $\epsilon, \epsilon' \in \g$.
In the case of a usual Hamiltonian action of $G$ on $M$, where $\delta_\epsilon$ is given by equation \eqref{CanonicalCase}, it is a well-known fact that the homomorphism condition \eqref{ActionLie} implies that the Poisson algebra of the charges $Q^a$ takes the form
\begin{equation*}
\{ Q^a, Q^b \} = \sum_c f^{ab}_{\hspace{8pt} c}\, Q^c + N^{ab}
\end{equation*}
of the Lie algebra relations in $\g$ up to central quantities $N^{ab}$, where $f^{ab}_{\hspace{8pt} c}$ are structure constants of $\g$ with respect to the basis $\{ T^a \}$, \emph{i.e.}
\begin{equation*}
[ T^a, T^b ] = \sum_c f^{ab}_{\hspace{8pt} c}\, T^c.
\end{equation*}
It is therefore natural to ask whether we can extract from equation \eqref{ActionLie} some informations on the Poisson bracket of $\Gamma$ with itself. One important step is to note that, from equation \eqref{MomentMap}, using the Jacobi and Leibniz identities on $\lbrace\cdot,\cdot\rbrace$, the action of $[\delta_\epsilon,\delta_{\epsilon'}]$ on any function $f$ takes the following rather simple form
\begin{equation}\label{ComDelta}
\left[ \delta_\epsilon,\delta_{\epsilon'} \right]f = \left\langle \epsilon\ti{1} \epsilon'\ti{2}, \Gamma\ti{1}^{-1} \Gamma\ti{2}^{-1} \left\lbrace \lbrace \Gamma\ti{1}, \Gamma\ti{2} \rbrace \Gamma\ti{1}^{-1} \Gamma\ti{2}^{-1}, f \right\rbrace \Gamma\ti{1} \Gamma\ti{2} \right\rangle\ti{12}.
\end{equation}
In order to treat the right hand side of equation \eqref{ActionLie}, we pass to the Drinfel'd double formulation recalled in section \ref{SubSec:Drinfeld}. Indeed, we can write
\begin{equation*}
\delta_{[\epsilon,\epsilon']}f = -\left\langle  [\iota(\epsilon),\iota(\epsilon')]_D \; \bigr| \, \iota^*(\Gamma)^{-1} \left\lbrace \iota^*(\Gamma), f \right\rbrace \right\rangle,
\end{equation*}
where, by abuse of notation, we still denote by $\iota^*$ the lift of $\iota^*:\g^* \hookrightarrow D\g$ to the group embedding $G^* \hookrightarrow DG$. Using the definition \eqref{RD} of $\Rc^D_{\1\2}$ we note that
\begin{equation*}
\iota(\epsilon')\ti{1} = \langle \iota(\epsilon')\ti{2} | \Rc^D\ti{12} \rangle\ti{2} \quad \text{and} \quad \iota(\epsilon)\ti{2} = -\langle \iota(\epsilon)\ti{1} | \Rc^D\ti{12} \rangle\ti{1}.
\end{equation*}
From these two equations and the fact that the pairing $\langle\cdot|\cdot\rangle$ is invariant with respect to the $[\cdot,\cdot]_D$ bracket (cf. Theorem \ref{thm: Drin double}), we obtain
\begin{equation*}
\delta_{[\epsilon,\epsilon']}f = -\frac{1}{2}\left\langle \iota(\epsilon)\ti{1} \iota(\epsilon')\ti{2} \, \Bigr|  \left[ \Rc^D\ti{12}, \iota^*(\Gamma)^{-1}\ti{1} \left\lbrace \iota^*(\Gamma)\ti{1}, f \right\rbrace + \iota^*(\Gamma)^{-1}\ti{2} \left\lbrace \iota^*(\Gamma)\ti{2}, f \right\rbrace \right]_D \right\rangle\ti{12}.
\end{equation*}
This expression can be rewritten as
\begin{equation}\label{DeltaCom}
\delta_{[\epsilon,\epsilon']}f = \frac{1}{2} \left\langle \iota(\epsilon)\ti{1} \iota(\epsilon')\ti{2} \, \Bigr| \, \iota^*(\Gamma)^{-1}\ti{1} \iota^*(\Gamma)^{-1}\ti{2} \left\lbrace \iota^*(\Gamma)\ti{1}\iota^*(\Gamma)\ti{2} \Rc^D\ti{12} \iota^*(\Gamma)^{-1}\ti{1} \iota^*(\Gamma)^{-1}\ti{2}, f \right\rbrace \iota^*(\Gamma)\ti{1}\iota^*(\Gamma)\ti{2} \right\rangle\ti{12}.
\end{equation}
Using the fact that the pairing $\langle\cdot|\cdot\rangle$ is non-degenerate between $\g$ and $\g^*$, by equating \eqref{ComDelta} and \eqref{DeltaCom} we arrive at
\begin{equation}\label{PBGammaP}
\left\lbrace \iota^*(\Gamma)\ti{1}, \iota^*(\Gamma)\ti{2} \right\rbrace \iota^*(\Gamma)^{-1}\ti{1} \iota^*(\Gamma)^{-1}\ti{2} = \frac{1}{2}
\iota^*(\Gamma)\ti{1}\iota^*(\Gamma)\ti{2} \Rc^D\ti{12} \iota^*(\Gamma)^{-1}\ti{1} \iota^*(\Gamma)^{-1}\ti{2} + P\ti{12}
\end{equation}
where the element $P\ti{12} \in D\g \otimes D\g$ is a central quantity for the Poisson bracket $\lbrace\cdot,\cdot\rbrace$. Let us study the properties that $P\ti{12}$ must fulfil. First of all, it must be skew-symmetric. Moreover, it should 
be such that the right hand side of \eqref{PBGammaP} lives in $\iota^*(\g^*)\otimes\iota^*(\g^*)$. It is a well-known 
consequence of the Adler-Kostant-Symes construction for Manin triples that, for any $y \in \iota^*(G^*)$, we have 
$y\ti{1}y\ti{2}\Rc^D\ti{12}y\ti{1}^{-1}y\ti{2}^{-1}-\Rc^D\ti{12} \in \iota^*(\g^*)\otimes\iota^*(\g^*)$. Thus, 
defining $N\ti{12}=P\ti{12}+\frac{1}{2}\Rc^D\ti{12}$, we can write
\begin{equation}\label{PBGamma}
\left\lbrace \iota^*(\Gamma)\ti{1}, \iota^*(\Gamma)\ti{2} \right\rbrace  =
-\frac{1}{2} \left[ \Rc^D\ti{12}, \iota^*(\Gamma)\ti{1}\iota^*(\Gamma)\ti{2} \right] + N\ti{12} 
\iota^*(\Gamma)\ti{1}\iota^*(\Gamma)\ti{2},
\end{equation}
with $N\ti{12} \in \iota^*(\g^*) \otimes \iota^*(\g^*)$ skew-symmetric. The last requirement on 
$N\ti{12}$ is that the Poisson bracket \eqref{PBGamma} must satisfy the Jacobi identity. Let us first remark that this is the case when $N\ti{12}=0$, as $\Rc^D\ti{12}$ verifies the mCYBE. We will see in the next sections why this case is of particular interest.

More generally, we recognise in \eqref{PBGamma} a quadratic algebra of $ad$-type, in the 
nomenclature of \cite{Freidel:1991jx,Freidel:1991jv}. In this case, a necessary and sufficient condition for the Jacobi identity to hold is that $\Rc^D\ti{12} - 2 N\ti{12}$ satisfies the mCYBE. In particular, this is the case if
\begin{equation*}
N\ti{12}=\frac{1}{2}\Rc^D\ti{12}-\frac{1}{2}\iota^*(C)\ti{1}^{-1}\iota^*(C)\ti{2}^{-1}\Rc^D\ti{12}\iota^*(C)\ti{1}\iota^*(C)\ti{2},
\end{equation*}
for some constant $C \in G^*$. For this $N\ti{12}$ we define $\tilde{\Gamma}=C\Gamma$. As noted in subsection \ref{subsec: namm}, $\tilde{\Gamma}$ is still a good non-abelian moment map. Moreover, the Poisson bracket of $\tilde{\Gamma}$ becomes
\begin{equation*}
\big\{ \iota^*(\tilde{\Gamma})\ti{1}, \iota^*(\tilde{\Gamma})\ti{2} \big\}  =
-\frac{1}{2} \big[ \Rc^D\ti{12}, \iota^*(\tilde{\Gamma})\ti{1}\iota^*(\tilde{\Gamma})\ti{2}\big].
\end{equation*}
Conversely, if the Poisson brackets of $\Gamma$ are of the form \eqref{PBGamma}, then the transformation \eqref{MomentMap} is a Lie algebra action of $\g$, \textit{i.e.} it satisfies \eqref{ActionLie}.

\section{Coboundary Poisson-Lie groups and $R$-matrices}
\label{Sec:Cobound}

One important class of Lie bialgebras are the so-called coboundary ones, which are given by $R$-matrices, solutions of the mCYBE. In this section, we recall their properties and apply the abstract result \eqref{PBGamma} of the previous section to this particular case.

\subsection[$R$-matrices, Sklyanin bracket and $\g_R$ dual algebra]{$R$-matrices, Sklyanin bracket and $\g_R$ dual algebra}
\label{SubSec:GR}

Let $\g$ be a Lie algebra and $R:\g\rightarrow\g$ a skew-symmetric linear map solution of the mCYBE on $\g$, namely
\begin{equation}\label{mCYBE}
[RX,RY]-R\bigl([RX,Y]+[X,RY]\bigr) = -c^2 [X,Y],
\end{equation}
for all $X,Y \in \g$ with $c=1$ (split case) or $c=i$ (non-split case). We define $R^\pm = R \pm c \, \Id$, and introduce the $R$-bracket
\begin{equation*}
[X,Y]_R = \gamma \big( [RX,Y]+[X,RY] \big) = \gamma \big( [R^\pm X, Y] + [X, R^\mp Y] \big),
\end{equation*}
with $\gamma$ a real constant. An important consequence of the mCYBE is that the vector space $\g$ equipped with the $R$-bracket is also a Lie algebra. We therefore have two Lie algebra structures on the vector space $\g$: the usual one $(\g,[\cdot,\cdot])$, that we shall still note $\g$ and
\begin{equation*}
\g_R = (\g,[\cdot,\cdot]_R).
\end{equation*}

This construction is related to Poisson-Lie groups. Suppose now that $\g$ is semisimple and let $\kappa$ denote its Killing form. Let $R$ be a skew-symmetric solution of the mCYBE on $\g$. We denote by $R\ti{12}\in\g\otimes\g$ its kernel with respect to $\kappa$. One can then define a Poisson-Lie structure on $G$ with the Sklyanin bracket
\begin{equation*}
\left\lbrace x\ti{1}, x\ti{2} \right\rbrace_G = \gamma \left[R\ti{12}, x\ti{1}x\ti{2} \right].
\end{equation*}
Since $\g$ is semisimple, its Killing form $\kappa$ is non-degenerate. This allows us to define a natural pairing $\pi$ between $\g$ and its dual $\g^*$ by considering, for any $X\in\g$, the linear form
\begin{align*}
\pi(X) : \g & \longrightarrow  \mathbb{R} \\
          Y & \longmapsto      \kappa(X,Y).
\end{align*}
As a vector space, $\g_R$ is equal to $\g$, so $\pi$ can be seen as a linear isomorphism from $\g_R$ to $\g^*$. The following lemma then gives a concrete realisation of the dual Lie algebra $\g^*$.

\begin{lemma}
Let $G$ be a Poisson-Lie group, with the Sklyanin bracket associated with a solution $R$ of the mCYBE on $\g$. Equip $\g^*$ with the Lie bracket \eqref{LieDual}. Then the map
\begin{equation*}
\pi : \g_R \longrightarrow \g^*
\end{equation*}
is a Lie algebra isomorphism.
\end{lemma}

\subsection[Real and complex doubles and the dual Lie algebras $\g_{DR}$ and $\g_\pm$]{Real and complex doubles and the dual Lie algebras $\g_{DR}$ and $\g_\pm$}
\label{SubSec:Doubles}

In this section, we study separately the split and non-split cases.

\paragraph{Split case.}
Define the real double of $\g$ as the Lie algebra direct sum
$\g \oplus \g$. We introduce the subspaces
\begin{equation*}
\gd = \bigl\lbrace (X,X), \, X\in\g \bigr\rbrace, \qquad
\g_{DR} = \bigl\lbrace (R^+X,R^-X), \, X\in\g \bigr\rbrace.
\end{equation*}
It is clear that, for any endomorphism $R$, $\gd$ and $\g_{DR}$ form a direct sum decomposition of $\g \oplus \g$. Moreover, $\gd$ is a Lie subalgebra of $\g \oplus \g$. One shows that, when $R$ is a split solution of the mCYBE, $\g_{DR}$ is also a Lie subalgebra of $\g \oplus \g$ isomorphic to $\g_R$. More precisely,

\begin{lemma}
If $R$ is a solution of the split mCYBE on $\g$, then
\begin{align*}
\Delta : \g_R & \longrightarrow  \g_{DR} \\
           X  & \longmapsto      \gamma(R^+X,R^-X)
\end{align*}
is a Lie algebra isomorphism, whose inverse is given for all $(X,Y) \in \g_{DR} \subset \g \oplus \g$ by
\begin{equation}\label{DeltaInv}
\Delta^{-1}(X,Y) = \frac{1}{2\gamma}(X-Y).
\end{equation}
\end{lemma}

We have obtained yet another realisation of $\g^*$, this time in the real double. Moreover the subalgebra $\gd$ is isomorphic to $\g$. Hence we have realised both $\g$ and $\g^*$ as subalgebras of the real double $\g \oplus \g$. In fact, by the following lemma the real double $\g \oplus \g$ itself is a realisation of the abstract Drinfel'd double $D\g=\g\oplus\g^*$ (cf. section \ref{SubSec:Drinfeld}).

\begin{lemma}\label{Lemma:PhiSplit}
If $R$ is a skew-symmetric solution of the split mCYBE on $\g$, then
\begin{align*}
\Phi : \quad D\g & \longrightarrow \g \oplus \g \\
           (X,\lambda)  & \longmapsto     (X,X) + \Delta\circ\pi^{-1}(\lambda)
\end{align*}
is a Lie algebra isomorphism, such that $\Phi\bigl(\iota(\g)\bigr)=\gd$ and $\Phi\bigl(\iota^*(\g^*)\bigr)=\g_{DR}$. Its inverse is given for every $(X,Y) \in \g \oplus \g$ by
\begin{equation*}
\Phi^{-1}(X,Y) = \frac{1}{2} \left( R^+Y-R^-X, \frac{1}{\gamma} \pi(X-Y) \right).
\end{equation*}
Moreover, $\Phi$ sends the pairing \eqref{PairingDg} on $D\g$ to the non-degenerate $\ad$-invariant bilinear form on $\g \oplus \g$ defined, for all $X_1, X_2, Y_1, Y_2 \in \g$, by
\begin{equation*}
\big\langle (X_1,Y_1) \big| (X_2,Y_2) \big\rangle = \frac{1}{2\gamma} \big( \kappa(X_1,X_2) - \kappa(Y_1,Y_2) \big).
\end{equation*}
\end{lemma}

\paragraph{Non-split case.}
 One can perform a similar analysis in the case of a non-split solution of the mCYBE ($c=i$). Here we introduce the complex double $\g^\C$ as the complexification of $\g$, namely
\begin{equation*}
\g^\C = \lbrace X+iY, \, X,Y \in \g \rbrace.
\end{equation*}
We define the complex conjugation relative to the real form $\g$ as
\begin{align*}
\theta : \qquad \g^\C & \longrightarrow \g^\C \\
         X+iY  & \longmapsto     X-iY.
\end{align*}
This is a semi-linear involutive automorphism of $\g^\C$ and $\g$ itself can be seen as a Lie subalgebra of $\g^\C$, viewed as a real Lie algebra. More precisely, $\g$ is the subalgebra of $\g^\C$ fixed by $\theta$ (see also appendix \ref{App:RealForms}).

 We introduce the subspaces
\begin{equation*}
\g_\pm = \lbrace R^\pm X, \, X\in\g \rbrace
\end{equation*}
of $\g^\C$. Note that $\g_\pm=\theta(\g_\mp)$. We have the vector space decompositions $\g^\C=\g\oplus\g_+ = \g\oplus\g_-$. Moreover, as a consequence of the mCYBE, $\g_\pm$ are Lie subalgebras of $\g^\C$ isomorphic to $\g_R$.

\begin{lemma}
If $R$ is a solution of the non-split mCYBE on $\g$, then
\begin{equation*}
\Delta_\pm = \gamma R^\pm : \g_R \longrightarrow \g_\pm
\end{equation*}
is a Lie algebra isomorphism, whose inverse is given for each $X \in \g_{\pm} \subset \g^\C$ by
\begin{equation}\label{DeltaInvNS}
\Delta_\pm^{-1}(X) = \pm\frac{X-\theta(X)}{2i\gamma}.
\end{equation}
\end{lemma}

 As in the split case, we realised $\g$ and $\g^*$ as subalgebras of $\g^\C$. Moreover, the complex double $\g^\C$ provides another realisation of the abstract Drinfel'd double $D\g$ by the following result.

\begin{lemma}\label{Lemma:PhiNonSplit}
If $R$ is a skew-symmetric solution of the non-split mCYBE on $\g$, then
\begin{align*}
\Phi_\pm : \quad D\g & \longrightarrow \g^\C \\
           (X,\lambda) & \longmapsto     X + \gamma R^\pm \circ\pi^{-1}(\lambda)
\end{align*}
is a Lie algebra isomorphism, such that $\Phi_\pm\bigl(\iota(\g)\bigr)=\g$ and $\Phi_\pm\bigl(\iota^*(\g^*)\bigr)=\g_\pm$. Its inverse is given for any $X \in \g^\C$ by
\begin{equation*}
\Phi_\pm^{-1}(X) = \frac{1}{2i} \left( R^\pm\big( \theta(X) \big) - R^\mp(X), \pm \frac{1}{\gamma} \pi \big( X - \theta(X) \big) \right).
\end{equation*}

Moreover, $\Phi_\pm$ sends the pairing \eqref{PairingDg} on $D\g$ to the non-degenerate $\ad$-invariant bilinear form on $\g^\C$ defined, for all $X, Y \in \g^\C$, by
\begin{equation*}
\left\langle X|Y \right\rangle = \pm\frac{1}{\gamma} \,\textup{Im} \big( \kappa(X,Y)\big).
\end{equation*}
\end{lemma}

\subsection[Poisson-Lie action of $G$: Semenov-Tian-Shansky brackets]{Poisson-Lie action of $G$: Semenov-Tian-Shansky brackets}
\label{SubSec:PL2STS}

In the previous subsections, we provided concrete realisations of both the dual Lie algebra $\g^\ast$ and the Drinfel'd double $D\g$ for (split and non-split) coboundary Poisson-Lie groups. By abuse of notation, we will denote by the same symbols the lift of these realisations to the dual group $G^\ast$ and the Drinfel'd double group $DG$. In section \ref{SubSec:PBGamma}, we found an abstract expression \eqref{PBGamma} for the Poisson bracket of the non-abelian moment map viewed in the Drinfel'd double. We will now investigate what this Poisson bracket becomes in the concrete realisations of $DG$.

\paragraph{Split case.}
The non-abelian moment map $\Gamma$ can be seen as a map to the group $G_R$ \textit{via} the Killing pairing $\pi$, namely
\begin{equation*}
\Gamma_R = \pi^{-1}(\Gamma) \in G_R,
\end{equation*}
and in turn as an element of the group $G_{DR}$ \textit{via} the morphism $\Delta$,
\begin{equation*}
(\Gamma^+,\Gamma^-) = \Delta(\Gamma_R) = \Delta\circ\pi^{-1}(\Gamma) \in G_{DR} \subset G\times G.
\end{equation*}
The real double $G \times G$ is related to the Drinfel'd double $DG$ by the morphism $\Phi$ (cf lemma \ref{Lemma:PhiSplit}). Let us remark here that
\begin{equation*}
\Phi\bigl(\iota^*(\Gamma)\bigr)=(\Gamma^+,\Gamma^-).
\end{equation*}
Under the action of $\Phi\ti{1}\Phi\ti{2}$, the Poisson bracket \eqref{PBGamma} then becomes
\begin{equation}\label{PBRealDouble}
\left\lbrace (\Gamma^+,\Gamma^-)\ti{1}, (\Gamma^+,\Gamma^-)\ti{2} \right\rbrace = -\frac{1}{2} \left[ \Phi\ti{1} \Phi\ti{2} \Rc^D\ti{12}, (\Gamma^+,\Gamma^-)\ti{1} (\Gamma^+,\Gamma^-)\ti{2} \right] + \tilde{N}\ti{12} (\Gamma^+,\Gamma^-)\ti{1} (\Gamma^+,\Gamma^-)\ti{2}
\end{equation}
with $\widetilde{N}\ti{12} = \Phi\ti{1}\Phi\ti{2} N\ti{12}$ the central quantity.

The objects in the above formula belong to $(\g \oplus \g) \otimes (\g \oplus \g)$. Such objects can be written as vectors with four $\g\otimes\g$-valued components as
\begin{equation*}
(X,Y) \otimes (X',Y') =
\begin{pmatrix}
X \otimes X' \\ X \otimes Y' \\ Y \otimes X' \\ Y \otimes Y'
\end{pmatrix}.
\end{equation*}
Let us now compute $\Phi\ti{1} \Phi\ti{2} \Rc^D\ti{12}$. We have $\Phi\circ\iota(T^a) = (T^a,T^a)$ and
\begin{equation*}
\Phi\circ\iota^*(T_a) = \Delta\circ\pi^{-1}(T_a) = \sum_b \kappa_{ab} \Delta(T^b) = \gamma\sum_b \kappa_{ab} (R^+T^b,R^-T^b)
\end{equation*}
where $\kappa_{ab}$ is the Killing form written in the basis $\lbrace T^a \rbrace$. We therefore find that
\begin{equation}\label{PhiRD}
\Phi\ti{1} \Phi\ti{2} \Rc^D\ti{12}
= -2\gamma \begin{pmatrix}
 R\ti{12} \\ R^+\ti{12} \\ R^-\ti{12} \\ R\ti{12}
 \end{pmatrix}.
\end{equation}
We define
\begin{equation*}
\widetilde{N}\ti{12} = \begin{pmatrix}
N^{++}\ti{12} \\ N^{+-}\ti{12} \\ N^{-+}\ti{12} \\ N^{--}\ti{12}
\end{pmatrix}.
\end{equation*}
The four components of the Poisson bracket \eqref{PBRealDouble} then read
\begin{subequations}
\begin{align}
\left\lbrace \Gamma^+\ti{1}, \Gamma^+\ti{2} \right\rbrace &= \gamma \left[ R\ti{12}, \Gamma^+\ti{1} \Gamma^+\ti{2} \right] + N^{++}\ti{12} \Gamma^+\ti{1} \Gamma^+\ti{2}, \\[5pt]
\left\lbrace \Gamma^+\ti{1}, \Gamma^-\ti{2} \right\rbrace &= \gamma \left[ R^+\ti{12}, \Gamma^+\ti{1} \Gamma^-\ti{2} \right] + N^{+-}\ti{12} \Gamma^+\ti{1} \Gamma^-\ti{2}, \\[5pt]
\left\lbrace \Gamma^-\ti{1}, \Gamma^+\ti{2} \right\rbrace &= \gamma \left[ R^-\ti{12}, \Gamma^-\ti{1} \Gamma^+\ti{2} \right] + N^{-+}\ti{12} \Gamma^-\ti{1} \Gamma^+\ti{2}, \\[5pt]
\left\lbrace \Gamma^-\ti{1}, \Gamma^-\ti{2} \right\rbrace &= \gamma \left[ R\ti{12}, \Gamma^-\ti{1} \Gamma^-\ti{2} \right] + N^{--}\ti{12} \Gamma^-\ti{1} \Gamma^-\ti{2}.
\end{align}
\end{subequations}
When the central quantities $N^{\pm\pm}\ti{12}$ and $N^{\pm\mp}\ti{12}$ vanish, these are the Semenov-Tian-Shansky brackets
\begin{subequations}\label{STSpb}
\begin{align}
\left\lbrace \Gamma^\pm\ti{1}, \Gamma^\pm\ti{2} \right\rbrace &= \gamma \left[ R\ti{12}, \Gamma^\pm\ti{1} \Gamma^\pm\ti{2} \right], \\
\left\lbrace \Gamma^\pm\ti{1}, \Gamma^\mp\ti{2} \right\rbrace &= \gamma \left[ R^\pm\ti{12}, \Gamma^\pm\ti{1} \Gamma^\mp\ti{2} \right].
\end{align}
\end{subequations}

Finally, let us emphasise that the transformation law \eqref{MomentMap} can be re-expressed in terms of the non-abelian moment map $(\Gamma^+,\Gamma^-)$ \textit{via} the morphism $\Delta\circ\pi^{-1}$, giving explicitly
\begin{equation} \label{ActionSplit}
\delta_\epsilon f = -\frac{1}{2\gamma} \kappa \Bigl( \epsilon, (\Gamma^+)^{-1}\left\lbrace \Gamma^+, f \right\rbrace - (\Gamma^-)^{-1}\left\lbrace \Gamma^-, f \right\rbrace \Bigr).
\end{equation}

\paragraph{Non-split case.}
From the non-abelian moment map $\Gamma_R=\pi^{-1}(\Gamma)$ seen in the group $G_R$, we can construct two different realisations of $\Gamma$ in the complex double $G^\C$:
\begin{equation*}
\Gamma^+ = \Delta_+ (\Gamma_R) \in G_+ \quad \text{and} \quad \Gamma^- = \Delta_- (\Gamma_R) \in G_-.
\end{equation*}
These are not independent. They are related by the semi-linear automorphism $\theta$ as $\Gamma^\pm = \theta(\Gamma^\mp)$. Note that
\begin{equation*}
\Phi_\pm\circ\iota^*(\Gamma) = \Gamma^\pm.
\end{equation*}
For any $\eta,\varepsilon \in \lbrace +,- \rbrace$, applying $\Phi_\eta\null\ti{1}\Phi_\varepsilon\null\ti{2}$ to the bracket \eqref{PBGamma}, we obtain
\begin{equation*}
\left\lbrace \Gamma^\eta\ti{1}, \Gamma^\varepsilon\ti{2} \right\rbrace = -\frac{1}{2} \left[ \Phi_\eta\null\ti{1}\Phi_\varepsilon\null\ti{2}\Rc^D\ti{12}, \Gamma^\eta\ti{1} \Gamma^\varepsilon\ti{2} \right] + N^{\eta\varepsilon}\ti{12} \Gamma^\eta\ti{1} \Gamma^\varepsilon\ti{2},
\end{equation*}
with the central quantities $N^{\eta\varepsilon}\ti{12} = \Phi_\eta\null\ti{1}\Phi_\varepsilon\null\ti{2} N\ti{12}$. We have
$\Phi_\pm\circ\iota(T^a)=T^a$
and
\begin{equation*}
\Phi_\pm\circ\iota^*(T_a)=\gamma R^\pm \bigl( \pi^{-1}(T_a) \bigr) =\gamma\sum_b \kappa_{ab} R^\pm T^b,
\end{equation*}
so that
\begin{equation*}
\Phi_\eta\null\ti{1}\Phi_\varepsilon\null\ti{2} \Rc^D\ti{12} = \gamma \bigl(R^\varepsilon\ti{21}-R^\eta\ti{12}\bigr)= -\gamma \bigl(R^{\eta}\ti{12}+R^{-\varepsilon}\ti{12} \bigr).
\end{equation*}
Thus, in the non-split case, the non-abelian moment maps $\Gamma^+$ and $\Gamma^-$ also satisfy, up to central quantities, the Semenov-Tian-Shansky Poisson brackets \eqref{STSpb}.\\

Applying $\Phi_\pm$ to equation \eqref{MomentMap} yields the transformation law in terms of the non-abelian moment map $\Gamma^\pm$ which reads
\begin{equation}\label{ActionNonSplit}
\delta_\epsilon f = \mp \frac{1}{\gamma} \, \textup{Im} \, \Bigl( \kappa \bigl( \epsilon, (\Gamma^\pm)^{-1}\left\lbrace \Gamma^\pm, f \right\rbrace  \bigr) \Bigr) =  -\frac{1}{2i\gamma} \kappa \Bigl( \epsilon, (\Gamma^+)^{-1}\left\lbrace \Gamma^+, f \right\rbrace - (\Gamma^-)^{-1}\left\lbrace \Gamma^-, f \right\rbrace \Bigr).
\end{equation}

\subsection[Poisson-Lie action of $G^*$: Sklyanin bracket on $U$]{Poisson-Lie action of $G^*$: Sklyanin bracket} \label{subsec: Sklyanin bracket}

In this section we consider the case of a coboundary Lie bialgebra specified by a split $R$-matrix. We have canonical isomorphisms $\g_{DR}^*\simeq\g^{**}\simeq\g$ of vector spaces. Moreover, the dual Lie group $G^\ast \simeq G_{DR}$ is a Poisson-Lie group when equipped with the Semenov-Tian-Shansky bracket
\begin{subequations}
\begin{align}
\left\lbrace x^\pm\ti{1}, x^\pm\ti{2} \right\rbrace &= \gamma \left[ R\ti{12}, x^\pm\ti{1} x^\pm\ti{2} \right], \\
\left\lbrace x^\pm\ti{1}, x^\mp\ti{2} \right\rbrace &= \gamma \left[ R^\pm\ti{12}, x^\pm\ti{1} x^\mp\ti{2} \right].
\end{align}
\end{subequations}
The induced Lie structure on $\g_{DR}^*$ is isomorphic to $\g$, so that the isomorphism $\g^{**}\simeq\g_{DR}^*\simeq\g$ also holds at the level of Lie algebras. Thus the Drinfel'd double $D\g^*$ of the dual Lie algebra $\g^\ast$ is isomorphic to the Drinfel'd double $D\g$ of the original Lie algebra $\g$.

As a consequence, the formalism developed in the previous sections can also be used to treat a Poisson-Lie action of the dual group $G^* \simeq G_{DR}$ on a Poisson manifold $M$. In this case, the non-abelian moment map is an application
\begin{equation*}
U : M \longrightarrow G.
\end{equation*}
We can see it as an element $\iota(U)$ of the Drinfel'd double $DG$. The results presented in section \ref{SubSec:PBGamma} still apply in this case and the Poisson bracket of $U$ is then given by
\begin{equation*}
\left\lbrace \iota(U)\ti{1}, \iota(U)\ti{2} \right\rbrace = -\frac{1}{2} \left[ \Rc^D\ti{12}, \iota(U)\ti{1}\iota(U)\ti{2} \right] + \iota_\1 \iota_\2 (M\ti{12}) \iota(U)\ti{1}\iota(U)\ti{2},
\end{equation*}
where $M\ti{12}$ is a central quantity valued in $\g \otimes \g$. Applying the morphism $\Phi$ to this equation, noting that $\Phi\bigl(\iota(U)\bigr)=(U,U)$ and recalling equation \eqref{PhiRD}, we obtain
\begin{equation}\label{PBU}
\left\lbrace U\ti{1}, U\ti{2} \right\rbrace = \gamma\left[ R\ti{12}, U\ti{1} U\ti{2} \right] + M\ti{12} U\ti{1}U\ti{2}.
\end{equation}
This is, up to central quantities, the Sklyanin bracket.

\section[Link with $q$-deformed algebras]{Link with $q$-deformed algebras}
\label{Sec:qAlgebra}

In this section we suppose that $\g$ is either the split real form or a non-split real form of the semisimple complexification $\g^\C$. The definitions and basic properties of semisimple complex Lie algebras are recalled in appendix \ref{App:SemiSimpleStruct} and those of (non-)split real forms in appendix \ref{App:RealForms}.

\subsection{Real forms and standard $R$-matrices}

Let $\pi_{\n_\pm}$ and $\pi_\h$ denote the projections with respect to the Cartan-Weyl decomposition $\g^\C=\h\oplus\n_+\oplus\n_-$.
We define the linear operator $R \in \End \g^\C$ by
\begin{equation*}
R = c \bigl( \pi_{\n_+} - \pi_{\n_-} \bigr)
\end{equation*}
with $c=1$ (split case) or $c=i$ (non-split case), which is a solution of the mCYBE on $\g^\C$. Moreover, as can be seen in the explicit bases \eqref{BasisSplit} and \eqref{BasisNonSplit} of the split and non-split real forms, this operator stabilises the real form $\g$. Thus, by restriction, it defines a solution of the mCYBE on $\g$, that we refer to as the standard $R$-matrix of $\g$.

 The kernel of $R$ with respect to the Killing form is
\begin{equation*}
R\ti{12} = c\sum_{\alpha > 0} \left(E_\alpha \otimes E_{-\alpha} - E_{-\alpha} \otimes E_\alpha \right).
\end{equation*}
Likewise, the kernel of $R^\pm$ is $R^\pm\ti{12}=R\ti{12}\pm c  \,C\ti{12}$, where the quadratic Casimir tensor is given
\begin{equation*}
C\ti{12} = \sum_{i=1}^l K_i \otimes K_i + \sum_{\alpha > 0} \left(E_\alpha \otimes E_{-\alpha} + E_{-\alpha} \otimes E_\alpha \right),
\end{equation*}
where $\{ K_i, i = 1, \ldots, l \}$ is an orthonormal basis of $\h$ with respect to the Killing form.

\subsection{Extraction of charges}

We saw in section \ref{SubSec:PL2STS} that, in the split case, the non-abelian moment map can be regarded as an element $(\Gamma^+,\Gamma^-)$ of $G_{DR} \subset G\times G$. In the non-split case, it can be represented as either $\Gamma^+ \in G_+ \subset G^\C$ or $\Gamma^- \in G_- \subset G^\C$ (with $\Gamma^+$ and $\Gamma^-$ related by $\Gamma^\pm=\theta(\Gamma^\mp)$, where $\theta$ is the semi-linear involutive automorphism of $G^\C$ defining the real subgroup $G$). Here we will treat the two cases together.

In the split (resp. non-split) case, the Lie algebras $R^\pm(\g)$ are the positive and negative Borel subalgebras of $\g$ (resp. $\g^\C$), with opposite Cartan parts. Therefore $\Gamma^+$ and $\Gamma^-$ are elements of the positive and negative Borel subgroups of $G$ (resp. $G^\C$), with Cartan parts inverses of one another. We choose to parametrise them as follows
\begin{equation}\label{DefDM}
\Gamma^+ = M^+D, \; \; \; \Gamma^-=D^{-1}M^-, \; \; \; M^\pm \in N_\pm, \; \; \; D \in H,
\end{equation}
where $H$ is the Cartan subgroup and $N_{\pm}$ the positive and negative unipotent subgroups of $G$.

We now extract scalar charges from $D$ and $M^\pm$. Starting with the Cartan part $D$, we choose a decomposition with respect to the basis of fundamental weights $P_i$, recalled in appendix \ref{App:BasesCartan},
\begin{equation}\label{DecompoD}
D = \exp\left(ic \gamma \sum_{i=1}^l Q^H_i P_i \right) = \prod_{i=1}^l \exp\left(i c\gamma Q^H_i P_i \right).
\end{equation}
The order of the product has no importance since the Cartan subgroup is abelian. We define
\begin{equation}\label{DefZ}
Z = i c \gamma\sum_{i=1}^l Q^H_i P_i \in \h,
\end{equation}
so that $D=\exp(Z)$.

The extraction of suitable charges from $M^+$ and $M^-$ is more involved. Let us fix a labelling $\beta_1,\ldots,\beta_n$ of the positive roots, where $n$ is the number of positive roots. We can parametrise $M^\pm$ by scalar charges $Q^E_\beta$ as
\begin{equation}\label{DecompoM}
M^\pm = \prod_{i=1}^n \exp\left( \pm ic\gamma A_{\pm\beta_i} Q^E_{\pm\beta_i} E_{\pm\beta_i} \right),
\end{equation}
where the $A_\beta$'s are normalisation constants to be fixed later. Define
\begin{equation} \label{DefUV}
u_{(i)} = \exp\left( ic\gamma A_{\beta_i} Q^E_{\beta_i} E_{\beta_i} \right) \;\;\;\; \text{ and } \;\;\;\; v_{(i)} = \exp\left( -ic\gamma A_{-\beta_i} Q^E_{-\beta_i} E_{-\beta_i} \right)
\end{equation}
so that we can write
\begin{equation}\label{DecompoUV}
M^+ = u_{(1)} \ldots u_{(n)}  \;\;\;\; \text{ and } \;\;\;\; M^- = v_{(1)} \ldots v_{(n)}.\vspace{8pt}
\end{equation}
Since the Lie groups $N_\pm$ are not abelian, these products depend on the choice of the ordering $\beta_1,\ldots,\beta_n$ of the positive roots. We choose an ordering such that
\begin{equation}\label{ConditionOrder}
\text{if } i < j \text{ and } \beta_i + \beta_j \text{ is a root, then } \beta_i + \beta_j = \beta_k \text{ with } i < k < j.
\end{equation}
Such an ordering can be constructed from the (partial) normal order described in~\cite{Delduc:2013fga}. We label the simple roots $\alpha_1, \ldots, \alpha_l$ in a way which is compatible with the ordering $\beta_1,\ldots,\beta_n$, \textit{i.e.} such that $\alpha_i=\beta_{k_i}$ with $1=k_1 \leq \ldots \leq k_l=n$.

\subsection[Semenov-Tian-Shansky brackets and $\U_q(\g)$ algebra]{Semenov-Tian-Shansky brackets and $\U_q(\g)$ algebra}
\label{SubSec:Uqg}

We will now start from the Semenov-Tian-Shansky brackets \eqref{STSpb} for the non-abelian moment map $\Gamma^\pm$ and extract from it the corresponding Poisson brackets between the charges $Q^H_i$ and $Q^E_\beta$, as defined above. We will make extensive use of two theorems for the extraction of Poisson brackets that we present in appendix \ref{App:ThmPB}. We shall denote by $\pi_k$ and $\pi_{-k}$ the projections onto $\C E_{\beta_k}$ and $\C E_{-\beta_k}$ with respect to the Cartan-Weyl decomposition
\begin{equation*}
\g = \bigoplus_{k=1}^n \left( \C E_{\beta_k} \oplus \C E_{-\beta_k} \right) \oplus \h.
\end{equation*}

\subsubsection[Poisson brackets of $D$ and $M^\pm$]{Poisson brackets of $D$ and $M^\pm$}

Consider the decomposition \eqref{DefDM} of $\Gamma^+$ and $\Gamma^-$. Using Theorem \ref{TheoremPB1}, we obtain
\begin{subequations}\label{PBDMM}
\begin{align}
\left\lbrace D\ti{1}, D\ti{2} \right\rbrace &= 0, \label{PBDD}\\
\left\lbrace D\ti{1}, M^\pm\ti{2} \right\rbrace &= \gamma D\ti{1} \left[H\ti{12}, M^\pm\ti{2} \right], \label{PBDM}\\
\left\lbrace M^+\ti{1}, M^-\ti{2} \right\rbrace &= \gamma \Bigl( D\ti{2} R^{++}\ti{12} D^{-1}\ti{2} M^+\ti{1}M^-\ti{2} - M^+\ti{1}M^-\ti{2} D^{-1}\ti{2}R^{++}\ti{12}D\ti{2} \Bigr), \label{PBMpMm}\\
\left\lbrace M^\pm\ti{1}, M^\pm\ti{2} \right\rbrace &= \gamma \Bigl( \left[ R\ti{12}, M^\pm\ti{1}M^\pm\ti{2} \right] \mp M^\pm\ti{1} H\ti{12} M^\pm\ti{2} \pm M^\pm\ti{2} H\ti{12} M^\pm\ti{1} \Bigr), \label{PBMM} 
\end{align}
\end{subequations}
where we have introduced
\begin{subequations}
\begin{align}
\label{H12 def} H\ti{12} &= c\sum_{i=1}^l K_i \otimes K_i,\\
R^{++}\ti{12} &= R^+\ti{12}-H\ti{12} = 2c\sum_{\alpha > 0} E_\alpha \otimes E_{-\alpha}, \\
R^{--}\ti{12} & = R^-\ti{12}+H\ti{12} = - 2c\sum_{\alpha > 0} E_{-\alpha} \otimes E_\alpha.
\end{align}
\end{subequations}
We also made use of the following identity, valid for any $h\in H$ and $\epsilon\in\lbrace \emptyset, +, ++, -, -- \rbrace$,
\begin{equation*}
h\ti{1}h\ti{2}R^\epsilon\ti{12}h\ti{1}^{-1}h\ti{2}^{-1}=R^\epsilon\ti{12}.
\end{equation*}
From the Poisson bracket \eqref{PBDD}, one simply finds
\begin{equation}\label{PBQHQH}
\left\lbrace Q^H_i, Q^H_j \right\rbrace = 0.
\end{equation}

\subsubsection[Poisson bracket between $Q^H_i$ and $Q^E_\beta$]{Poisson bracket between $Q^H_i$ and $Q^E_\beta$}
\label{SubSubSec:QHQE}

The partial Casimir tensor \eqref{H12 def} on the Cartan subalgebra can be expressed in terms of the dual bases of weights $P_i$ and co-roots $\alpha_i^\vee$ (cf appendix \ref{App:BasesCartan}) as
\begin{equation*}
H\ti{12} = c\sum_{i=1}^l P_i \otimes \alpha_i^\vee.
\end{equation*}
This allows us to extract the Poisson bracket between $Q^H_i$ and $M^\pm$ by projecting equation \eqref{PBDM} onto $P_i$ in the first tensor factor, namely
\begin{equation*}
i \lbrace Q^H_i, M^\pm \rbrace = \left[ \alpha_i^\vee, M^\pm \right].
\end{equation*}

 We will now treat the bracket with $M^+$, the case of $M^-$ being similar. We introduce
\begin{equation*}
w_{(k)} = u_{(k)} \ldots u_{(n)},
\end{equation*}
such that $M^+=w_{(1)}$ and $w_{(k)}=u_{(k)}w_{(k+1)}$. Using this decomposition and Theorem \ref{TheoremPB1}, one shows by induction on $k$ that, for every $k\in\lbrace1,\ldots,n\rbrace$, we have
\begin{subequations}
\begin{align}
i u_{(k)}^{-1} \lbrace Q^H_i, u_{(k)} \rbrace &= u_{(k)}^{-1} \alpha_i^\vee u_{(k)} - \alpha_i^\vee, \label{PBQHu}\\
i \lbrace Q^H_i, w_{(k)} \rbrace w_{(k)}^{-1} &= \alpha_i^\vee - w_{(k)} \alpha_i^\vee w_{(k)}^{-1}. 
\end{align}
\end{subequations}
This induction relies on the fact that for any $k$, the adjoint action of $w_{(k+1)}$ on $\alpha_i^\vee$ only creates nilpotent generators $E_\gamma$ corresponding to roots of the form $\gamma=a_{k+1}\beta_{k+1} + \ldots + a_n\beta_n$, with $a_{k+1},\ldots,a_n \in \N$. One can show from the ordering condition \eqref{ConditionOrder} that these roots are always strictly superior to the root $\beta_k$, which allows to perform the projection needed in Theorem \ref{TheoremPB1}. Using the definition \eqref{DefUV} of $u_{(k)}$, equation \eqref{PBQHu} becomes
\begin{equation*}
i \lbrace Q^H_i, Q^E_{\beta_k} \rbrace = \beta_k(\alpha_i^\vee) Q^E_{\beta_k}.
\end{equation*}
Applying the same method to the Poisson bracket with $M^-$, we find that this equation holds for any root $\beta$, positive or negative. In the case of a simple root (or its opposite) $\beta=\pm\alpha_j$, we have $\beta(\alpha_i^\vee)=\pm\alpha_j(\alpha_i^\vee)=\pm A_{ij}$ (cf appendix \ref{App:BasesCartan}). We therefore obtain
\begin{equation}\label{PBQEQH}
i \lbrace Q^H_i, Q^E_{\pm\alpha_j} \rbrace = \pm A_{ij} Q^E_{\pm\alpha_j}.
\end{equation}

\subsubsection[Poisson bracket between $Q^E_{\alpha_i}$ and $Q^E_{-\alpha_j}$]{Poisson bracket between $Q^E_{\alpha_i}$ and $Q^E_{-\alpha_j}$}

Fixing two simple roots $\alpha_i$ and $\alpha_j$, we want to compute the Poisson bracket between $Q^E_{\alpha_i}$ and $Q^E_{-\alpha_j}$. Recall that $\alpha_i=\beta_{k_i}$ and $\alpha_j=\beta_{k_j}$. Considering the decomposition \eqref{DecompoUV} of $M^\pm$, we need to extract the Poisson bracket of $u_{(k_i)}$ with $v_{(k_j)}$. Define
\begin{align*}
x &=u_{(1)} \ldots u_{(k_i)}, & \tilde{x} &= v_{(1)} \ldots v_{(k_j)},\\
y &=u_{(k_i+1)} \ldots u_{(n)}, & \tilde{y} &= v_{(k_j+1)} \ldots v_{(n)}.
\end{align*}
By Theorem \ref{TheoremPB2}, applied on both tensor factors, we may write
\begin{equation*}
u_{(k_i)\,}^{-1}\null\ti{1} v_{(k_j)\,}^{-1}\null\ti{2} \left\lbrace u_{(k_i)\,}\null\ti{1}, v_{(k_j)\,}\null\ti{2} \right\rbrace = \pi_{k_i} \otimes \pi_{-k_j} \bigl( \Pc\ti{12} \bigr),
\end{equation*}
where $\Pc\ti{12} = x^{-1}\ti{1}\tilde{x}^{-1}\ti{2} \left\lbrace M^+\ti{1}, M^-\ti{2} \right\rbrace y^{-1}\ti{1} \tilde{y}^{-1}\ti{2}$. On the other hand, from equation \eqref{PBMpMm} we find
\begin{equation*}
\Pc\ti{12} = \gamma \Bigl( x^{-1}\ti{1}\tilde{x}^{-1}\ti{2} D\ti{2} R^{++}\ti{12} D^{-1}\ti{2} x\ti{1}\tilde{x}\ti{2} - y\ti{1} \tilde{y}\ti{2} D^{-1}\ti{2}R^{++}\ti{12}D\ti{2} y^{-1}\ti{1} \tilde{y}^{-1}\ti{2} \Bigr).
\end{equation*}
Recalling from \eqref{DefZ} that $D=\exp(Z)$ with $Z\in\h$, we have
\begin{equation*}
D^{\pm 1}\ti{2}R^{++}\ti{12}D^{\mp 1}\ti{2} = 2c \sum_{\alpha > 0} \exp\bigl(\mp \alpha(Z)\bigr) E_\alpha \otimes E_{-\alpha},
\end{equation*}
so that
\begin{equation*}
\Pc\ti{12} = 2c \gamma \sum_{\alpha > 0}  \Bigl( \exp\bigl(- \alpha(Z)\bigr) \left( x^{-1} E_\alpha x \right) \otimes \left( \tilde{x}^{-1} E_{-\alpha} \tilde{x} \right) - \exp\bigl(\alpha(Z)\bigr) \left( y E_\alpha y^{-1} \right) \otimes \left( \tilde{y} E_{-\alpha} \tilde{y}^{-1} \right) \Bigr).
\end{equation*}
The adjoint action of any $E_\beta$ (appearing in $x$ or $y$) on $E_\alpha$ cannot create the simple root generator $E_{\alpha_i}$ and similarly for $E_{-\alpha_j}$ on the second space. It follows that
\begin{equation*}
\pi_{k_i} \otimes \pi_{-k_j} \bigl( \Pc\ti{12} \bigr) = 2c \gamma \delta_{ij} \Bigl( \exp\bigl(- \alpha_i(Z)\bigr) - \exp\bigl(\alpha_i(Z)\bigr) \Bigr)  E_{\alpha_i} \otimes E_{-\alpha_j}.
\end{equation*}
Yet, by definition \eqref{DefUV} of the $u_{(k)}$'s and $v_{(k)}$'s, we find
\begin{equation*}
\pi_{k_i} \otimes \pi_{-k_j} \bigl( \Pc\ti{12} \bigr) = u_{(k_i)\,}^{-1}\null\ti{1} v_{(k_j)\,}^{-1}\null\ti{2} \left\lbrace u_{(k_i)\,}\null\ti{1}, v_{(k_j)\,}\null\ti{2} \right\rbrace = c^2 \gamma^2 A_{\alpha_i} A_{-\alpha_j} \left\lbrace Q^E_{\alpha_i}, Q^E_{-\alpha_j} \right\rbrace E_{\alpha_i} \otimes E_{-\alpha_j},
\end{equation*}
so that
\begin{equation*}
i \big\{ Q^E_{\alpha_i}, Q^E_{-\alpha_j} \big\} = \frac{2i}{c\gamma A_{\alpha_i} A_{-\alpha_j}} \delta_{ij} \Bigl( \exp\bigl(-\alpha_i(Z)\bigr) - \exp\bigl(\alpha_i(Z)\bigr) \Bigr).
\end{equation*}
From equation \eqref{DefZ}, one has (cf. appendix \ref{App:BasesCartan})
\begin{equation*}
\alpha_i(Z) = ic\gamma \sum_{k=1}^l Q_k^H \alpha_i(P_k) = i c\gamma \sum_{k=1}^l Q_k^H d_i \delta_{ik} = i c\gamma d_i Q^H_i.
\end{equation*}
Introducing the deformation parameter
\begin{equation}\label{Defq}
q=e^{-ic\gamma},
\end{equation}
we therefore have
\begin{equation*}
i \big\{ Q^E_{\alpha_i}, Q^E_{-\alpha_j} \big\} = \frac{2i}{\gamma c A_{\alpha_i} A_{-\alpha_j}} \delta_{ij} \Bigl( q^{d_iQ^H_i}-q^{-d_iQ^H_i} \Bigr).
\end{equation*}
Finally, if we fix the normalisation $A_{\pm \alpha}$ for simple roots as
\begin{equation}\label{Aalphai}
A_{\pm\alpha_i} = \left( \frac{4 \sinh(i c \gamma d_i)}{ic \gamma} \right)^{\frac{1}{2}},
\end{equation}
then we may rewrite the above Poisson brackets as
\begin{equation}\label{PBQEpQEm}
i \big\{ Q^E_{\alpha_i}, Q^E_{-\alpha_j} \big\} = \delta_{ij} \frac{q^{d_iQ^H_i}-q^{-d_iQ^H_i}}{q^{d_i}-q^{-d_i}}.
\end{equation}

\subsubsection[$q$-Poisson-Serre relations]{$q$-Poisson-Serre relations}
\label{SubSubSec:qSerre}

The Poisson brackets \eqref{PBQHQH}, \eqref{PBQEQH} and \eqref{PBQEpQEm} obtained so far are part of the defining relations of the semiclassical limit $\U_q(\g)$ of the quantum group $U_{\widehat q}(\g)$ with $\widehat q = q^\hbar$, as given in \cite{Delduc:2013fga}. The complete set of relations characterising $\U_q(\g)$ also includes the so-called $q$-Poisson-Serre relations. The purpose of the present subsection is to derive these from the Poisson bracket \eqref{PBMM}. We will only treat the case of positive roots, the negative one being handled similarly.

\paragraph{Poisson brackets of $Q^E_{\alpha_i}$ with $M^+$.}
Let us fix a simple root $\alpha_i$. We recall that $\alpha_i=\beta_{k_i}$, so that $Q^E_{\alpha_i}$ is to be extracted from $u_{(k_i)}$. Introduce
\begin{align*}
x &= u_{(1)} \ldots u_{(k_i)}, \\
y &= u_{(k_i+1)} \ldots u_{(n)},
\end{align*}
so that $M^+=xy$. By Theorem \ref{TheoremPB2} we have
\begin{equation*}
\left(u_{(k_i)}\right)\ti{1}^{-1} \left\lbrace u_{(k_i)}\null\ti{1}, M^+\ti{2} \right\rbrace = \left(\pi_{k_i}\right)\ti{1} \bigl( \Pc\ti{12} \bigr),
\end{equation*}
where $\Pc\ti{12} = x^{-1}\ti{1} \left\lbrace M^+\ti{1}, M^+\ti{2} \right\rbrace y^{-1}\ti{1}$.
On the other hand, from \eqref{PBMM} we have
\begin{equation*}
\Pc\ti{12} = \gamma \Pc^R\ti{12} + \gamma\Pc^H\ti{12} 
\end{equation*}
with
\begin{align*}
\Pc^R\ti{12} &= x^{-1}\ti{1} R^{++}\ti{12} x\ti{1} M^+\ti{2} - M^+\ti{2} y\ti{1} R^{++}\ti{12} y^{-1}\ti{1}, \\
\Pc^H\ti{12} &= \left( x^{-1}\ti{1} H\ti{12} x\ti{1} - y\ti{1}H\ti{12}y\ti{1}^{-1} \right) M^+\ti{2} + M^+\ti{2} \left( x^{-1}\ti{1} H\ti{12} x\ti{1} - y\ti{1}H\ti{12}y\ti{1}^{-1} \right).
\end{align*}
By writing \eqref{H12 def} as $H\ti{12}=\displaystyle c\sum_{j=1}^l \omega_j^\vee\otimes H_j$ (cf. appendix \ref{App:BasesCartan}), these can be rewritten
\begin{align*}
\Pc^R\ti{12} &= 2c\sum_{\alpha > 0} \bigl( x^{-1} E_\alpha x \bigr) \otimes \bigl( E_{-\alpha} M^+ \bigr) - 2c\sum_{\alpha > 0} \bigl( yE_\alpha y^{-1} \bigr) \otimes \bigl( M^+ E_{-\alpha} \bigr), \\
\Pc^H\ti{12} &= c\sum_{j=1}^l \bigl(x^{-1} \omega_j^\vee x - y \omega_j^\vee y^{-1} \bigr) \otimes \bigl( H_j M^+ + M^+ H_j \bigr).
\end{align*}
The adjoint action of any $E_\beta$ (appearing in $x$ or $y$) on $E_\alpha$ cannot create the simple root generator $E_{\alpha_i}$. Thus, we have
\begin{equation*}
\left(\pi_{k_i}\right)\ti{1} \bigl( \Pc^R\ti{12} \bigr) = 2c \, E_{\alpha_i} \otimes \bigl( E_{-\alpha_i} M^+ -  M^+ E_{-\alpha_i} \bigr).
\end{equation*}
In the same way, in the adjoint actions of $E_\beta$'s from $x$ or $y$ on $\omega^\vee_j$, only a unique adjoint action of $E_{\alpha_i}$, coming from $u_{(k_i)}$ in $x$, can create the simple root generator $E_{\alpha_i}$. Therefore
\begin{equation*}
\pi_{k_i} \bigl(x^{-1} \omega_j^\vee x - y \omega_j^\vee y^{-1} \bigr) = -ic\gamma A_{\alpha_i} Q^E_{\alpha_i} \ad_{E_{\alpha_i}} \bigl( \omega_j^\vee\bigr) = ic\gamma A_{\alpha_i} Q^E_{\alpha_i} \delta_{ij} E_{\alpha_i},
\end{equation*}
and hence
\begin{equation*}
\left(\pi_{k_i}\right)\ti{1} \bigl( \Pc^H\ti{12} \bigr) = ic^2 \gamma A_{\alpha_i} Q^E_{\alpha_i} E_{\alpha_i} \otimes \bigl( H_i M^+ + M^+ H_i \bigr).
\end{equation*}
Putting together all the above we arrive at
\begin{equation*}
\left(u_{(k_i)}\right)\ti{1}^{-1} \left\lbrace u_{(k_i)}\null\ti{1}, M^+\ti{2} \right\rbrace = c\gamma E_{\alpha_i} \otimes \Bigl( 2\bigl[ E_{-\alpha_i}, M^+ \bigr] + ic\gamma A_{\alpha_i} Q^E_{\alpha_i} \bigl( H_i M^+ + M^+ H_i \bigr) \Bigr).
\end{equation*}
Yet, by definition of $u_{(k_i)}$ in \eqref{DefUV} we have
\begin{equation*}
\left(u_{(k_i)}\right)\ti{1}^{-1} \left\lbrace u_{(k_i)}\null\ti{1}, M^+\ti{2} \right\rbrace = ic\gamma A_{\alpha_i} E_{\alpha_i} \otimes \left\lbrace Q^E_{\alpha_i}, M^+ \right\rbrace,
\end{equation*}
and hence
\begin{equation}\label{PBQMp}
i A_{\alpha_i} \left\lbrace Q^E_{\alpha_i}, M^+ \right\rbrace = 2\bigl[ E_{-\alpha_i}, M^+ \bigr] + ic\gamma A_{\alpha_i} Q^E_{\alpha_i} \bigl( H_i M^+ + M^+ H_i \bigr).
\end{equation}

\paragraph{$\alpha_i$-string through $\alpha_j$.}
Let us now consider another simple root $\alpha_j$. We suppose here that $\alpha_i > \alpha_j$. The $\alpha_i$-string through $\alpha_j$ is then contained between $\alpha_j$ and $\alpha_i$. Specifically, we have
\begin{equation*}
\alpha_j < \alpha_j + \alpha_i < \ldots < \alpha_j - A_{ij} \alpha_i <  \alpha_i,
\end{equation*}
with $A$ the Cartan matrix of $\g$ (cf. appendix \ref{App:BasesCartan}). Let $r \in \lbrace 0, \ldots, -A_{ij} \rbrace$ and $p \in \lbrace 1,\ldots,n \rbrace$ be such that
\begin{equation*}
\beta_p = \alpha_j + r \alpha_i,
\end{equation*}
We define
\begin{align*}
x &= u_{(1)} \ldots u_{(p)}, \\
y &= u_{(p+1)} \ldots u_{(n)},
\end{align*}
and $\Qc = x^{-1} \left\lbrace Q_{\alpha_i}^E, M^+ \right\rbrace y^{-1}$. By Theorem \ref{TheoremPB2}, we have
\begin{equation} \label{pip up}
u_{(p)}^{-1} \left\lbrace Q^E_{\alpha_i}, u_{(p)} \right\rbrace = \pi_p (\Qc).
\end{equation}
On the other hand, from the Poisson bracket \eqref{PBQMp} we get
\begin{equation}\label{Qc}
i A_{\alpha_i} \Qc = 2 \bigl( x^{-1} E_{-\alpha_i} x - y E_{-\alpha_i} y^{-1} \bigr) + i c\gamma A_{\alpha_i} Q^E_{\alpha_i} \bigl( x^{-1} H_i x + y H_i y^{-1} \bigr).
\end{equation}

The projection onto $E_{\beta_p}$ of the terms involving $H_i$ on the right hand side of \eqref{Qc} can be computed as follows. We note that $y H_i y^{-1}$ contains nilpotent generators $E_\beta$ with $\beta$ a sum of roots superior to $\beta_p$, which therefore cannot be $\beta_p$. In the same way, the adjoint action of $x$ on $H_i$ creates nilpotent generators $E_\beta$ with $\beta$ a sum of roots inferior or equal to $\beta_p$. Such $\beta$ can be either strictly inferior to $\beta_p$ or $\beta_p$ itself. Therefore the only way to have $E_{\beta_p}$ in $x^{-1}H_i x$ is by the simple adjoint action on $H_i$ of the generator $E_{\beta_p}$ (appearing in $u_{(p)}$). Thus, we have
\begin{equation}\label{ProjH}
\pi_p \bigl( x^{-1} H_i x + y H_i y^{-1} \bigr) = -ic\gamma A_{\beta_p} Q^E_{\beta_p} \ad_{E_{\beta_p}}\bigl(H_i\bigr) = ic\gamma A_{\beta_p} Q^E_{\beta_p} (\alpha_i,\beta_p) E_{\beta_p}.
\end{equation}

Next, consider the term $x^{-1} E_{-\alpha_i} x - y E_{-\alpha_i} y^{-1}$ on the right hand side of \eqref{Qc}. It is composed of generators $E_\beta$, with $\beta=\gamma-\alpha_i$ and $\gamma$ a sum of roots either all inferior or equal to $\beta_p$ (for $x$) or all superior (for $y$). We want to project this on $E_{\beta_p}$. Yet, having $\beta=\beta_p$ requires $\gamma=\beta_p+\alpha_i$. As $\beta_p < \alpha_i$, this means that $\beta_p < \gamma$, hence $\gamma$ comes from the adjoint action of $y$. To be more precise, $yE_{-\alpha_i}y^{-1}$ is composed of elements of the form (up to prefactors)
\begin{equation*}
\ad^{a_{p+1}}_{E_{\beta_{p+1}}} \ldots \ad^{a_n}_{E_{\beta_n}} \bigl( E_{-\alpha_i} \bigr), \; \; \; \text{ with } a_{p+1},\ldots,a_n \in \N.
\end{equation*}
Such a term is proportional to $E_{\gamma-\alpha_i}$, with $\gamma=a_{p+1}\beta_{p+1}+\ldots+a_n\beta_n$. In order to get $E_{\beta_p}$, one must have $\gamma=\alpha_i+\beta_p=\alpha_j+(r+1)\alpha_i$. Therefore, we want to solve
\begin{equation*}
a_{p+1}\beta_{p+1}+\ldots+a_n\beta_n = \alpha_j+(r+1)\alpha_i,
\end{equation*}
with $a_{p+1},\ldots,a_n$ non-negative integers. If a root $\beta_q$, for $q>p$, contains a simple root $\alpha_k$ different from $\alpha_i$ and $\alpha_j$, it is clear from the equation above that $a_q$ must be zero, as $\alpha_k$ does not appear in the right hand side of the equation.\\
\indent Moreover, the only roots superior to $\beta_p$ and containing only $\alpha_i$ and $\alpha_j$ as simple roots are $\alpha_j+(r+1)\alpha_i, \alpha_j+(r+2)\alpha_i, \ldots, \alpha_j-A_{ij}\alpha_i$ and $\alpha_i$. The only way that a non-negative integer linear combination of these roots can give $\alpha_j+(r+1)\alpha_i$ is if all the coefficients are zero except for that of the root $\alpha_j+(r+1)\alpha_i$ itself. Thus, the projection of $x^{-1} E_{-\alpha_i} x - y E_{-\alpha_i} y^{-1}$ onto $E_{\beta_p}$ comes from the simple adjoint action of $E_{\alpha_j+(r+1)\alpha_i}$ on $E_{-\alpha_i}$ (if $\alpha_j+(r+1)\alpha_i$ is a root). Hence
\begin{equation}\label{ProjF}
\begin{split}
\pi_p \bigl( x^{-1} E_{-\alpha_i} x - y E_{-\alpha_i} y^{-1} \bigr) &= - ic\gamma A_{\alpha_j+(r+1)\alpha_i} Q^E_{\alpha_j+(r+1)\alpha_i} \bigl[ E_{\alpha_j+(r+1)\alpha_i}, E_{-\alpha_i} \bigr] \\
 &= - ic\gamma A_{\alpha_j+(r+1)\alpha_i} Q^E_{\alpha_j+(r+1)\alpha_i} N_{\beta_p,\alpha_i} E_{\beta_p},
\end{split}
\end{equation}
if $\alpha_j+(r+1)\alpha_i$ is a root, and is zero otherwise.

Applying $\pi_p$ to \eqref{Qc} and using the results \eqref{ProjH} and \eqref{ProjF} gives
\begin{equation*}
iA_{\alpha_i} \pi_p(\Qc) = -2ic\gamma A_{\alpha_j+(r+1)\alpha_i} N_{\beta_p,\alpha_i} Q^E_{\alpha_j+(r+1)\alpha_i} E_{\beta_p} + (ic\gamma)^2 A_{\alpha_i} A_{\beta_p} (\alpha_i,\beta_p) Q^E_{\alpha_i} Q^E_{\beta_p} E_{\beta_p}.
\end{equation*}
Yet from \eqref{pip up} together with the definition of $u_{(p)}$ in \eqref{DefUV} we have
\begin{equation*}
\pi_p (\Qc) = u_{(p)}^{-1} \big\{ Q^E_{\alpha_i}, u_{(p)} \big\} = ic\gamma A_{\beta_p} \big\{ Q^E_{\alpha_i}, Q^E_{\beta_p} \big\} E_{\beta_p},
\end{equation*}
and hence
\begin{equation*}
\big\{ Q^E_{\alpha_i}, Q^E_{\beta_p} \big\} = \frac{ A_{\alpha_j+(r+1)\alpha_i}}{A_{\alpha_i} A_{\beta_p}} 2 i N_{\beta_p,\alpha_i} Q^E_{\alpha_j+(r+1)\alpha_i} + c \gamma (\alpha_i,\beta_p) Q^E_{\alpha_i} Q^E_{\beta_p}.
\end{equation*}
We define the $q$-bracket of two charges associated with the positive roots $\alpha$ and $\beta$ as
\begin{equation*}
\big\{ Q^E_\alpha, Q^E_\beta \big\}_q = \left\lbrace Q^E_\alpha, Q^E_\beta \right\rbrace + c\gamma (\alpha,\beta) Q^E_\alpha Q^E_\beta.
\end{equation*}
Moreover, if we fix the normalisation constant $A_\alpha$ for $\alpha$ in the $\alpha_i$-string through $\alpha_j$ as
\begin{equation*}
A_{\alpha_j+r\alpha_i} = A_{\alpha_j}A_{\alpha_i}^r,
\end{equation*}
then we deduce that
\begin{equation*}
\big\{ Q^E_{\alpha_j+r\alpha_i}, Q^E_{\alpha_i} \big\}_q =   2i N_{\alpha_i,\alpha_j+r\alpha_i} Q^E_{\alpha_j+(r+1)\alpha_i},
\end{equation*}
if $\alpha_j+(r+1)\alpha_i$ is a root and is zero otherwise.

By induction, we get the $q$-Poisson-Serre relation
\begin{equation}
\bigl\lbrace \bigl\lbrace \ldots \bigl\lbrace Q_{\alpha_j}^E, \underbrace{Q_{\alpha	_i}^E \bigr\rbrace_q, \ldots Q_{\alpha_i}^E \bigr\rbrace_q, Q_{\alpha_i}^E}_{1-A_{ij} \text{ times}} \bigr\rbrace_q = 0.
\end{equation}
One can treat the case $\alpha_i < \alpha_j$ in a similar way. For that, one needs to use a slightly different version of Theorem \ref{TheoremPB2}, involving the quantity
$\left( u_{(1)} \ldots u_{(p-1)} \right)^{-1} \left\lbrace Q^E_{\alpha_i}, M^+ \right\rbrace \left( u_{(p)} \ldots u_{(n)} \right)^{-1}$ instead of
$\left( u_{(1)} \ldots u_{(p)} \right)^{-1} \left\lbrace Q^E_{\alpha_i}, M^+ \right\rbrace \left( u_{(p+1)} \ldots u_{(n)} \right)^{-1}$. This yields the $q$-Poisson-Serre relation
\begin{equation}
\bigl\lbrace \underbrace{Q_{\alpha_i}^E, \bigl\lbrace Q_{\alpha_i}^E, \ldots, \bigl\lbrace Q^E_{\alpha_i}}_{1-A_{ij} \text{ times}}, Q^E_{\alpha_j} \bigr\rbrace_q \ldots \bigr\rbrace_q \bigr\rbrace_q = 0.
\end{equation}

Applying the same method as above to the Poisson bracket in \eqref{PBMM} involving $M^-$, one finds that the charges $Q^E_{-\alpha_i}$ also verifiy $q$-Poisson-Serre relations, but with respect to the deformed bracket $\lbrace\cdot,\cdot\rbrace_{q^{-1}}$, defined for two negative roots $\alpha$ and $\beta$ as
\begin{equation*}
\big\{ Q_\alpha^E, Q_\beta^E \big\}_{q^{-1}} = \big\{ Q^E_\alpha, Q^E_\beta \big\} - c\gamma (\alpha,\beta) Q^E_\alpha Q^E_\beta.
\end{equation*}

\subsubsection{Reality conditions}
\label{SubSubSec:RealCond}

The Poisson brackets \eqref{PBQHQH}, \eqref{PBQEQH} and \eqref{PBQEpQEm}, together with the $q$-Poisson-Serre relations stated above are the defining Poisson bracket relations of the semiclassical limit $\U_q(\g)$ of the quantum group $U_{\widehat q}(\g)$ with $\widehat q = q^\hbar$. It only remains to check that the required reality conditions are verified by the charges $Q^H_i$ and $Q^E_\alpha$. We shall address this question in the present subsection, first in the split case and then in the non-split one.

\paragraph{Split case.}
When $c=1$, the deformation parameter \eqref{Defq} becomes
\begin{equation*}
q=e^{-i\gamma},
\end{equation*}
so that $|q|=1$, \textit{i.e.} $q$ is a phase. Now the moment map $(\Gamma^+,\Gamma^-)$ takes values in the real double $G \times G$, therefore the reality condition is simply
\begin{equation*}
\theta(\Gamma^\pm)=\Gamma^\pm, 
\end{equation*}
with $\theta$ the split semi-linear automorphism described in appendix \ref{App:RealForms} and lifted to the complexified group $G^\C$. As $\theta$ stabilises the Cartan and the unipotent subgroups and since the decomposition \eqref{DefDM} is unique, one has
\begin{equation*}
\theta(D)=D \; \; \; \; \text{ and } \; \; \; \; \theta(M^\pm)=M^\pm.
\end{equation*}
We recall (cf. appendix \ref{App:RealForms}) that in the split case $\theta(P_i)=P_i$ for $i\in\lbrace 1,\ldots,l \rbrace$ (since $P_i$ is a real linear combination of the $H_j$) and $\theta(E_\alpha)=E_\alpha$ for any root $\alpha$. Considering the extraction of charges \eqref{DecompoD} and \eqref{DecompoM} with $c=1$, the above reality condition gives
\begin{align*}
\overline{Q^H_i} &= -Q^H_i, \\
\overline{A_{\pm\alpha_i} Q^E_{\pm\alpha_i}} &= -A_{\pm\alpha_i}Q^E_{\pm\alpha_i}.
\end{align*}
The normalisation constants $A_{\pm\alpha_i}$ are given by equation \eqref{Aalphai}, which in the split case reads
\begin{equation*}
A_{\pm\alpha_i} = \left( \frac{4 \sin(\gamma d_i)}{\gamma} \right)^{\frac{1}{2}}.
\end{equation*}
We will restrict attention to the case where
\begin{equation*}
-\pi \leq \gamma d_i \leq \pi,
\end{equation*}
for any $i$, so that the $A_{\pm\alpha_i}$ are real numbers. As a result, the reality conditions are simply
\begin{equation}
|q|=1, \; \; \; \; \overline{Q^H_i} = -Q^H_i \; \; \; \; \text{and} \; \; \; \; \overline{Q^E_{\pm\alpha_i}} = -Q^E_{\pm\alpha_i}.
\end{equation}
These are the reality conditions for the split real form $\U_q(\g)$, which correspond precisely to the semiclassical counterpart of the reality conditions on $U_{\widehat q}(\g)$ as given in \cite{Ruegg:1993eq}.

\paragraph{Non-split case.}
For $c=i$, the deformation parameter \eqref{Defq} now reads
\begin{equation*}
q=e^\gamma,
\end{equation*}
which is a real number. As explained in subsection \ref{SubSec:PL2STS}, the two moment maps $\Gamma^+$ and $\Gamma^-$ are not independent. They are related by the reality condition
\begin{equation*}
\theta(\Gamma^\pm)=\Gamma^\mp,
\end{equation*}
where $\theta$ is the non-split semi-linear automorphism described in appendix \ref{App:RealForms}, lifted to the complexified group $G^\C$. Since $\theta$ stabilises the Cartan subgroup $H$ but exchanges the unipotent subgroups $N_\pm$, applying $\theta$ to the decomposition \eqref{DefDM} we get
\begin{align*}
\theta(\Gamma^+) &= \theta(M^+)\theta(D)=\underbrace{\theta(D)}_{\in H} \underbrace{\theta(D)^{-1}\theta(M^+)\theta(D)}_{\in N_-},\\
\theta(\Gamma^-) &= \theta(D)^{-1}\theta(M^-) = \underbrace{\theta(D)^{-1}\theta(M^-)\theta(D)}_{\in N_+} \underbrace{\theta(D)^{-1}}_{\in H},
\end{align*}
where we used the fact that an adjoint action of a Cartan element on a element of $N_\pm$ is still in $N_\pm$. Equating $\theta(\Gamma^+)$ with $\Gamma^-=D^{-1}M^-$ and $\theta(\Gamma^-)$ with $\Gamma^+=M^+D$ we obtain
\begin{equation} \label{theta D M}
\theta(D) = D^{-1} \; \; \; \; \text{ and } \; \; \; \; \theta(M^\pm) = D^{-1} M^\mp D.
\end{equation}
Recall (cf. appendix \ref{App:RealForms}) that $\theta(P_i)=-P_i$. Using the extraction of Cartan charges \eqref{DecompoD}, with $c=i$, we find
\begin{equation*}
\overline{Q^H_i}=Q^H_i.
\end{equation*}
From the decomposition \eqref{DecompoM}, we have
\begin{equation*}
D^{-1}M^\pm D = \prod_{k=1}^n \exp\left( \mp \gamma A_{\pm\beta_k} Q^E_{\pm\beta_k} D^{-1} E_{\pm\beta_k} D \right).
\end{equation*}
Moreover, since $D=\exp(Z)$ with $Z$ defined in \eqref{DefZ},
\begin{equation*}
D^{-1} E_{\pm\beta_k} D = \exp\bigl(\mp \beta_k(Z)\bigr) E_{\pm\beta_k} = \exp\bigl(\pm \gamma \sum_{j=1}^l Q_j^H \beta_k(P_j)\bigr)E_{\pm\beta_k}.
\end{equation*}
In particular, for $k=k_i$, \textit{i.e.} for $\beta_k$ the simple root $\alpha_i$, using \eqref{AlphaP} we get
\begin{equation*}
D^{-1} E_{\pm\alpha_i} D = q^{\pm d_i Q_i^H}E_{\pm\alpha_i}.
\end{equation*}
The term corresponding to the simple root $\alpha_i$ in $D^{-1}M^\mp D$ therefore reads
\begin{equation*}
\exp \Bigl( \pm \gamma A_{\mp\alpha_i} q^{\mp d_i Q^H_i} Q^E_{\mp\alpha_i} E_{\mp\alpha_i} \Bigr).
\end{equation*}
Yet we have $\theta(E_{\pm\alpha_i})=-\lambda_i E_{\mp\alpha_i}$, so that the corresponding term in $\theta(M^\pm)$ is
\begin{equation*}
\exp \Bigl( \pm \gamma \overline{A_{\pm\alpha_i} Q^E_{\pm\alpha_i}} \lambda_i E_{\mp\alpha_i} \Bigr).
\end{equation*}
It now follows from the second equality in \eqref{theta D M} that
\begin{equation*}
\overline{A_{\pm\alpha_i} Q^E_{\pm\alpha_i}} = \lambda_i q^{\mp d_i Q^H_i} A_{\mp\alpha_i} Q^E_{\mp\alpha_i}.
\end{equation*}
The normalisation constants $A_{\pm\alpha_i}$ are given by \eqref{Aalphai}, which for $c=i$ take the form
\begin{equation*}
A_{\pm\alpha_i} = \left( \frac{4 \sinh(\gamma d_i)}{\gamma} \right)^{\frac{1}{2}}
\end{equation*}
and are therefore real numbers. Hence, the reality conditions are
\begin{equation}
q\in\mathbb{R}, \; \; \; \; \overline{Q^H_i}=Q^H_i \; \; \; \; \text{and} \; \; \; \; \overline{ Q^E_{\pm\alpha_i}}=\lambda_i q^{\mp d_i Q^H_i} Q^E_{\mp\alpha_i}.
\end{equation}
According to~\cite{Ruegg:1993eq}, these are the reality conditions of the non-split real form $\U_q(\g)$.

\subsection[Sklyanin bracket and $\U_q(\g^*)$ algebra]{Sklyanin bracket and $\U_q(\g^*)$ algebra}

As in subsection \ref{subsec: Sklyanin bracket}, in what follows we consider only the split case. We start from the Poisson bracket \eqref{PBU} with the central quantity $\widetilde{M}\ti{12}$ set to zero. In other words, $U$ satisfies the Sklyanin Poisson bracket
\begin{equation}\label{PBU2}
\left\lbrace U\ti{1}, U\ti{2} \right\rbrace = \gamma\left[ R\ti{12}, U\ti{1} U\ti{2} \right].
\end{equation}
Let us decompose $U$ as
\begin{equation*}
U=M^-DM^+,
\end{equation*}
with $D \in H$ and $M^\pm \in N_\pm$. Using Theorem \ref{TheoremPB1}, we can extract the Poisson brackets between $D$, $M^+$ and $M^-$ from equation \eqref{PBU2} to find
\begin{subequations}
\begin{align}
\left\lbrace D\ti{1}, D\ti{2} \right\rbrace &= 0,\\
\left\lbrace D\ti{1}, M^\pm\ti{2} \right\rbrace &= \pm\gamma D\ti{1} \left[H\ti{12}, M^\pm\ti{2} \right], \label{PBDM2}\\
\left\lbrace M^+\ti{1}, M^-\ti{2} \right\rbrace &= 0, \label{PBMpMm2}\\\left\lbrace M^\pm\ti{1}, M^\pm\ti{2} \right\rbrace &= \gamma \Bigl( \left[ R\ti{12}, M^\pm\ti{1}M^\pm\ti{2} \right] \mp M^\pm\ti{1} H\ti{12} M^\pm\ti{2} \pm M^\pm\ti{2} H\ti{12} M^\pm\ti{1} \Bigr). \label{PBMM2}
\end{align}
\end{subequations}
We notice that these Poisson brackets are very similar to \eqref{PBDMM}, the main difference being that $M^+$ and $M^-$ Poisson commute in the present case.

The methods of subsection \ref{SubSec:Uqg} can be applied to this case. For each positive root $\alpha$, we extract the positive nilpotent charge $Q^E_\alpha$ from $M^+$ and negative nilpotent charge $Q^E_{-\alpha}$ from $M^-$ as we did before using the decomposition \eqref{DecompoM}. Likewise, we extract Cartan charges $Q^H_i$ from $D$ as we did in equation \eqref{DecompoD}.

From equation \eqref{PBMM2}, following the same procedure outlined in subsection \ref{SubSubSec:qSerre}, we find that the nilpotent charges $Q^E_\alpha$ and $Q^E_{-\alpha}$ satisfy the $q$-Poisson-Serre relations. In other words, these charges span nilpotent $q$-deformed Poisson algebras $\U_q(\n_+)$ and $\U_q(\n_-)$. Moreover, it is clear from \eqref{PBMpMm2} that elements from these two algebras Poisson commute.

Similarly, we can apply the methods of subsection \ref{SubSubSec:QHQE} to the Poisson bracket \eqref{PBDM2}. Doing so, we find that, for any positive root $\alpha$,
\begin{equation*}
i \lbrace Q^H_i, Q^E_\alpha \rbrace = \alpha(\alpha_i^\vee) Q^E_\alpha \; \; \; \; \text{ and } \; \; \; \; i \lbrace Q^H_i, Q^E_{-\alpha} \rbrace = \alpha(\alpha_i^\vee) Q^E_{-\alpha}.
\end{equation*}
This implies that the Cartan charges $Q^H_i$ for $i=1,\ldots,n$ together with the charges $Q^E_\alpha$ for $\alpha > 0$ span a $q$-deformed positive Borel algebra $\U_q(\b_+)$. In the same way, the charges $-Q^H_i$ for $i=1,\ldots,n$ together with the charges $Q^E_{-\alpha}$ for $\alpha > 0$ span a $q$-deformed negative Borel algebra $\U_q(\b_-)$. The combination of all the charges $Q^H_i$ for $i=1,\ldots,n$ and $Q^E_\alpha$ for all roots $\alpha$ therefore span a $q$-deformed Poisson algebra which we could call $\U_q(\g_{DR})$.

Since $U$ takes value in the split real form $G$ we have $\theta(U)=U$, with $\theta$ the split semi-linear automorphism of appendix \ref{App:RealForms}. Moreover, since $\theta$ stabilises the subgroups $H$ and $N_\pm$, we deduce that $\theta(D)=D$ and $\theta(M^\pm)=M^\pm$. The reality conditions are then identical to those of the split case in the subsection \ref{SubSubSec:RealCond}, so that
\begin{equation}
|q|=1, \; \; \; \; \; \overline{Q^H_i}=-Q^H_i \; \; \; \; \text{and} \; \; \; \; \; \overline{Q^E_{\pm\alpha}}=-Q^E_{\pm\alpha}.
\end{equation}

\section{Application to Yang-Baxter type models} \label{sec: YB models}

\subsection{Yang-Baxter type models}

In this section we will apply the formalism of Poisson-Lie groups and non-abelian moment maps to discuss the symmetries of Yang-Baxter type models. The latter can be defined as the result of applying a general procedure for constructing integrable deformations of a broad family of integrable models. We briefly recall the construction below, referring to \cite{Vicedo:2015pna} for details.

We consider models whose classical integrable structure is described by a Lax matrix $\L(\lambda, \sigma)$ which is rational in the spectral parameter $\lambda \in \C$ and taking values in the space of $\g^\C$-valued fields on the real line (parameterised by $\sigma$). We 
require the Poisson bracket of the Lax matrix with itself to take the 
following general non-ultralocal form \cite{Maillet:1985fn,Maillet:1985ek}
\begin{align*}
\{ \L_\1(\lambda, \s), \L_\2(\mu, \s') \} &= \big[ \mathscr R_{\1\2}(\lambda, \mu), \L_\1(\lambda, \s) \big] \delta_{\s \s'} - \big[ \mathscr R_{\2\1}(\mu, \lambda), \L_\2(\mu, \s) \big] \delta_{\s \s'}\\
&\qquad\qquad\qquad\qquad - \big( \mathscr R_{\1\2}(\lambda, \mu) + \mathscr R_{\2\1}(\mu, \lambda) \big) \delta'_{\s\s'},
\end{align*}
where $\delta_{\s\s'}$ denotes the Dirac $\delta$-distribution and $\delta'_{\s\s'} = \partial_\s \delta_{\s\s'}$.
Here $\mathscr R_{\1\2}(\lambda, \mu)$ is a non-skew-symmetric $\g^\C \otimes \g^\C$-valued rational function of $\lambda$ and $\mu$ satisfying the classical Yang-Baxter equation with spectral parameter
\begin{equation*}
\big[ \mathscr R_{\1\2}(\lambda, \mu), \mathscr R_{\1\3}(\lambda, \nu) \big] + \big[ \mathscr R_{\1\2}(\lambda, \mu), \mathscr R_{\2\3}(\mu, \nu) \big] + \big[ \mathscr R_{\3\2}(\nu, \mu), \mathscr R_{\1\3}(\lambda, \nu) \big] = 0.
\end{equation*}
We will assume that its leading order behaviour in the limit $\lambda \to \mu$ is of the following form
\begin{equation*}
\mathscr R_{\1\2}(\lambda, \mu) = \frac{C_{\1\2}}{\mu - \lambda} \varphi(\mu)^{-1} + \mathcal O\big( (\lambda - \mu)^0 \big),
\end{equation*}
where $\varphi(\lambda)$ is called the twist function, which is rational in $\lambda$ and plays a central role in the construction. For certain models the Lax matrix $\L(\lambda, \sigma)$ may be twisted by some automorphism of the complex Lie algebra $\g^\C$, \emph{e.g.} for symmetric space $\sigma$-models, and the construction of Yang-Baxter type deformations applies equally to these. However, to keep the presentation concise we will not discuss the subtleties relating to this case. We note however that the main equations \eqref{PB gg gX XX} and \eqref{Lg} below remain valid.

In the original undeformed integrable model, we assume that the twist function $\varphi(\lambda)$ has a double pole at some point $\lambda_0 \in \mathbb{R}$. One can show that, associated with this double pole is a pair of fields $g$ valued in $G$ and $X$ valued in $\g$ with the following Poisson brackets
\begin{subequations} \label{PB gg gX XX}
\begin{align}
\left\lbrace g\ti{1}(\s), g\ti{2}(\s') \right\rbrace & = 0, \\
\left\lbrace X\ti{1}(\s), g\ti{2}(\s') \right\rbrace & = g\ti{2}(\s) \, C\ti{12} \,  \delta_{\s\s'}, \\
\left\lbrace X\ti{1}(\s), X\ti{2}(\s') \right\rbrace & = -\left[ C\ti{12}, X\ti{2}(\s) \right] \delta_{\s\s'}.\label{PBXX}
\end{align}
\end{subequations}
To construct a Yang-Baxter type model we begin by modifying the twist function $\varphi(\lambda)$ of the undeformed model, by deforming its double pole at $\lambda_0$ to a pair of simple poles which we denote $\lambda_\pm$, while keeping all other poles and zeroes fixed. In order to preserve certain reality conditions of the Lax matrix, we should take either $\lambda_\pm$ both real or $\lambda_+$ and $\lambda_-$ complex conjugate of one another. We refer to these two cases as the real and complex branches respectively.
Finally, we require also that the deformed twist function $\varphi(\lambda)$ has opposite residues at the simple poles $\lambda_+$ and $\lambda_-$, which allows us to define a single real deformation parameter $\gamma \in \mathbb{R}$ by
\begin{equation*}
\frac{c}{2 \gamma} = \res_{\lambda_-} \varphi(\lambda) d\lambda = - \res_{\lambda_+} \varphi(\lambda) d\lambda,
\end{equation*}
with $c = 1$ in the real branch and $c = i$ in the complex branch.

The Yang-Baxter type model may now be defined as follows. We fix a choice of $R$-matrix $R : \g \rightarrow \g$ which is a split or non-split solution of the mCYBE on $\g$, depending on whether the twist function $\varphi(\lambda)$ has been deformed in the real or complex branch. Keeping the same Lax matrix $\L(\lambda, \sigma)$ as in the original undeformed model, one can show \cite{Delduc:2013fga,Vicedo:2015pna} that the $G$-valued field $g$ and $\g$-valued field $X$ such that
\begin{equation}\label{Lg}
\Lc^g(\lambda_\pm,\s) = \gamma R^\pm X(\s),
\end{equation}
where $\Lc^g(\lambda, \sigma)$ denotes the gauge transformation of $\Lc(\lambda, \sigma)$ by $g$, satisfy the same fundamental Poisson brackets \eqref{PB gg gX XX} as in the undeformed case. In other words, the underlying phase space is still the same after the deformation process, but the expression of the Lax matrix in terms of the new fields $g$ and $X$, and hence also the Hamiltonian, is now modified.
In the following analysis, we shall take the equations \eqref{PB gg gX XX} and \eqref{Lg} as our starting point for studying the symmetries in Yang-Baxter type models.

For any $\sigma > \sigma'$ we define the transition matrix
\begin{equation}\label{Transition}
T^g(\lambda_\pm \, ; \s,\s')= \Pexp \left( -\int_{\s'}^\s \dd\rho \, \Lc^g(\lambda_\pm,\rho) \right).
\end{equation}
The main properties of path-ordered exponentials we shall need are recalled in appendix \ref{App:PathExp}. In particular, we have
\vspace{-12pt}\begin{subequations}\label{DiffPExp}
\begin{align}
\bigl( \p_\s T^g(\lambda_\pm \, ;\s,\s') \bigr) T^g(\lambda_\pm \, ;\s,\s')^{-1} &= -\Lc^g(\lambda_\pm, \s), \\
T^g(\lambda_\pm \, ;\s,\s')^{-1} \bigl( \p_{\s'} T^g(\lambda_\pm \, ;\s,\s') \bigr) &= \Lc^g(\lambda_\pm, \s').
\end{align}
\end{subequations}
Moreover, we suppose that the different currents of the models (such as $X$ and $g^{-1}\p_\s g$) vanish when $\s$ tends to $\pm \infty$. Therefore, since $\Lc^g(\lambda_\pm, \sigma)$ is the spatial component of a flat connection,
\begin{equation*}
T^g(\lambda_\pm) = T^g \bigl(\lambda_\pm\, ; +\infty,-\infty \bigr)
\end{equation*}
is conserved (cf. appendix \ref{App:PathExp}).

\subsection[Poisson brackets of $T^g(\lambda_\pm)$]{Poisson brackets of $T^g(\lambda_\pm)$}

In this subsection, we will compute the Poisson bracket of $T^g(\lambda_\pm)$ with itself and with $T^g(\lambda_\mp)$. In general, when a Lax matrix obeys Poisson brackets of the $r/s$-type, the Poisson bracket of its path-ordered exponential is ill-defined, due to the presence of non-ultralocal terms.

However, in our present case, as $\Lc^g(\lambda_\pm, \sigma)$ depends only on the field $X$, the Poisson bracket of $\Lc^g(\lambda_\pm, \sigma)$ with itself is ultralocal. More precisely, starting from \eqref{PBXX} and \eqref{Lg} we find
\begin{equation}\label{PBLgLg}
\left\lbrace \Lc^g_\1(\lambda_\pm,\s_1), \Lc^g_\2(\lambda_\pm,\s_2) \right\rbrace = \gamma \left[  \Lc^g_\1(\lambda_\pm,\s_1) + \Lc^g_\2(\lambda_\pm,\s_2), R\ti{12} \right] \delta_{\s_1\s_2},
\end{equation}
where we have used the mCYBE equation \eqref{mCYBE}. According to equation \eqref{dPExp} we have
\begin{align*}
\left\lbrace T^g_\1(\lambda_\pm), T^g_\2(\lambda_\pm) \right\rbrace & = \displaystyle \int_{-\infty}^{+\infty} \dd\s_1 \int_{-\infty}^{+\infty} \dd\s_2 \; \,  T^g_\1(\lambda_\pm \, ; +\infty,\s_1) T^g_\2(\lambda_\pm \, ; +\infty,\s_2)\\
 & \qquad \times \left\lbrace \Lc^g_\1(\lambda_\pm,\s_1), \Lc^g_\2(\lambda_\pm,\s_2) \right\rbrace T^g_\1 (\lambda_\pm \, ; \s_1,-\infty) T^g_\2 (\lambda_\pm \, ; \s_2,-\infty).
\end{align*}
Inserting \eqref{PBLgLg} into the latter and integrating the $\delta$-distribution, we obtain an expression for $\left\lbrace T^g_\1(\lambda_\pm), T^g_\2(\lambda_\pm) \right\rbrace$ as a single integral. Using the differential equations \eqref{DiffPExp} to re-express $\Lc^g(\lambda_\pm, \sigma)$ in terms of the transition matrices \eqref{Transition}, we recognise the total derivative of a product and obtain, after integration,
\begin{equation*}
\left\lbrace T^g_\1(\lambda_\pm), T^g_\2(\lambda_\pm) \right\rbrace = \gamma \left[ R\ti{12}, T^g_\1(\lambda_\pm) T^g_\2(\lambda_\pm) \right].
\end{equation*}
Similarly, we can compute the Poisson brackets between $T^g(\lambda_+)$ and $T^g(\lambda_-)$ to find
\begin{equation*}
\left\lbrace T^g_\1(\lambda_+), T^g_\2(\lambda_-) \right\rbrace = \gamma \left[ R^+\ti{12}, T^g_\1(\lambda_+) T^g_\2(\lambda_-) \right].
\end{equation*}
In conclusion, $T^g(\lambda_+)$ and $T^g(\lambda_-)$ satisfy the Semenov-Tian-Shansky Poisson brackets \eqref{STSpb}.

\subsection[Poisson-Lie $G$-symmetry]{Poisson-Lie $G$-symmetry}

For the remainder of this section we shall restrict attention to the non-split case. The treatment of the split case is completely analogous.

\subsubsection{The non-abelian moment map}

According to \eqref{Lg}, $\Lc^g(\lambda_\pm, \sigma)$ take values in the subalgebras $\g_{\pm}$ of the complex double $\g^\C$ (cf. subsection \ref{SubSec:Doubles}). Hence the path-ordered exponentials $T^g(\lambda_\pm \, ; \s,\s')$ belong to the subgroups $G_\pm$, which are realisations of the dual group $G^*$.
Moreover, we proved in the previous subsection that $T^g(\lambda_\pm)$ satisfies the Semenov-Tian-Shansky bracket. It follows from subsection \ref{SubSec:PBGamma} that $T^g(\lambda_\pm)$ has the right Poisson brackets for being the non-abelian moment map of a Poisson-Lie action of $G$. In the notations of the previous sections, we therefore consider
\begin{equation}\label{GammapmYB}
\Gamma^\pm = T^g(\lambda_\pm).
\end{equation}
as the two realisations of a non-abelian moment map in $G_\pm$, embedded in the complex double $G^\C$. It is natural to also look for the expression of this non-abelian moment map in the other realisation of the dual group $G^*$, namely the group $G_R$ described in subsection \ref{SubSec:GR}. This is given by $\Gamma_R=\Delta_\pm^{-1} \bigl( T^g(\lambda_\pm) \bigr)$, with $\Delta_\pm : G_R \rightarrow G_\pm$ the automorphisms described in subsection \ref{SubSec:Doubles}. In order to evaluate this explicitly we note that \eqref{Lg} can be written as $\Lc^g(\lambda_\pm, \sigma) = \Delta_\pm X(\sigma)$. Therefore, according to equation \eqref{AutoPExp}, the non-abelian moment map seen in $G_R$ simply reads
\begin{equation}\label{GammaRYB}
\Gamma_R = \Pexp_{G_R} \left( - \int_{-\infty}^{\infty} \dd\s \; X(\s) \right).
\end{equation}
In this expression, $X(\s)$ is seen as an element of $\g_R$ and the path-ordered exponential is taken in the group $G_R$.

\subsubsection[Transformation law of $g$ and $X$]{Transformation law of $g$ and $X$}

As motivated in the previous subsection, we consider the Poisson-Lie action of $G$ generated by the non-abelian moment map $T^g(\lambda_\pm) \in G_\pm$. According to equation \eqref{ActionNonSplit}, the infinitesimal form of this action is given by
\begin{equation}\label{TransfoTg}
\delta_\epsilon f = -\frac{1}{2i\gamma} \kappa \Bigl( \epsilon, T^g(\lambda_+)^{-1}\left\lbrace T^g(\lambda_+), f \right\rbrace - T^g(\lambda_-)^{-1}\left\lbrace T^g(\lambda_-), f \right\rbrace \Bigr) = -\kappa \bigl( \epsilon, \Gamma_R^{-1} \left\lbrace\Gamma_R,f\right\rbrace \bigr).
\end{equation}

In the undeformed case $\gamma=0$, the group $G_R$ is abelian and, from the expression \eqref{GammaRYB} of the non-abelian moment map $\Gamma_R$, the transformation \eqref{TransfoTg} becomes the usual Hamiltonian action with moment map $Q=\int_{-\infty}^{+\infty} \dd\s \; X(\s)$, namely
\begin{equation*}
\delta_\epsilon f = \kappa \bigl(\epsilon,\lbrace Q, f \rbrace \bigr).
\end{equation*}
This corresponds to the action of $G$ by right multiplication on $g$ since
\begin{equation}\label{TransfogXNonDef}
\delta_\epsilon g(\s) = g(\s)\epsilon \; \; \; \; \text{ and } \; \; \; \; \delta_\epsilon X(\s) = [X(\s), \epsilon].
\end{equation}
We will see in the rest of this subsection that, for $\gamma \neq 0$, the Poisson-Lie action generated by $T^g(\lambda_\pm)$ is still a right multiplication of $g$, but with a more complicated parameter.

Since the Poisson bracket of $\Lc^g(\lambda_\pm, \sigma)$ with the fields $g$ and $X$ is ultralocal, we can compute the Poisson brackets of $T^g(\lambda_\pm)$ with $g$ and $X$ using equation \eqref{dPExp}, without the need for any regularisation. We find
\begin{align*}
\left\lbrace T^g_\1(\lambda_\pm), g_\2(\s) \right\rbrace &= -\gamma T^g_\1(\lambda_\pm \, ; +\infty, \sigma)\, g\ti{2}(\s)\, R^\pm\ti{12}\, T^g_\1(\lambda_\pm \, ; \sigma, -\infty), \\
\left\lbrace T^g_\1(\lambda_\pm), X_\2(\s) \right\rbrace &= - \gamma T^g_\1(\lambda_\pm \, ; +\infty, \sigma)\, \left[ X\ti{2}(\s), R^\pm\ti{12} \right] T^g_\1(\lambda_\pm \, ; \sigma, -\infty).
\end{align*}
Inserting these expressions into \eqref{TransfoTg}, we obtain the transformation law of $g$ and $X$,
\begin{equation}\label{TransfogX}
\delta_\epsilon g(\s) = g(\s)K(\s) \; \; \; \; \text{ and } \; \; \; \; \delta_\epsilon X(\s) = [X(\s), K(\s)],
\end{equation}
where we have defined
\begin{align}\label{DefK}
K(\s) &= \frac{1}{2i} R^+\Bigl( T^g(\lambda_- \, ; \sigma,-\infty) \, \epsilon \, T^g(\lambda_- \, ; \sigma,-\infty)^{-1} \Bigr) \notag\\
&\qquad\qquad - \frac{1}{2i} R^-\Bigl( T^g(\lambda_+ \, ; \sigma,-\infty) \, \epsilon \, T^g(\lambda_+ \, ; \sigma,-\infty)^{-1} \Bigr).
\end{align}
We note that this transformation law has the same structure as the undeformed one \eqref{TransfogXNonDef} but with $\epsilon$ replaced by a more complicated (and non-constant) expression $K(\s)$. In particular, this field is non-local, as it contains $T^g(\lambda_\pm \, ; \sigma,-\infty)$. Since $T^g(\lambda_\pm \, ; \sigma,-\infty)$ becomes equal to the identity when $\gamma=0$, we see that $K$ turns back into $\epsilon$ in the undeformed case.

According to the paragraph following equation \eqref{MomentMap}, the transformation \eqref{TransfogX} must preserve the Poisson brackets on $(g,X)$ if $\epsilon$ possesses a Poisson bracket with itself, coming from the linearisation of the Sklyanin Poisson bracket on $G$, namely
\begin{equation}\label{SklyaninAlg}
\{ \epsilon\ti{1}, \epsilon\ti{2} \} = \gamma [ R\ti{12}, \epsilon\ti{1} + \epsilon\ti{2} ].
\end{equation}
For coherence, one can check this directly from the expression \eqref{DefK} of $K(\s)$. This (slightly long) computation involves some algebraic manipulations to simplify the expressions, in particular the identity
\begin{equation*}
\Ad_{T^g(\lambda_\pm)} \circ R^\pm = R^\pm \circ \Ad^{G_R}_{\Gamma_R},
\end{equation*}
which is a consequence of equation \eqref{AutoPExp}, applied to the automorphism $\Delta_\pm$.

The transformation law \eqref{TransfogX} may seem complicated because of the non-local expression \eqref{DefK} for $K(\s)$. However, it can be re-interpreted in a simpler way by introducing the more adapted variables \cite{Klimcik:2002zj,Delduc:2013fga,Vicedo:2015pna}
\begin{equation*}
\Psi_\pm(\s) = g(\s) x_\pm(\s), \; \; \; \; \text{ with } \; \; \; \; x_\pm(\s) = T^g(\lambda_\pm \, ; \s,-\infty).
\end{equation*}
In terms of these, the quantity \eqref{DefK} may be then rewritten as
\begin{equation}\label{DefK2}
K(\s) = \frac{1}{2i} R^+\Bigl( x_-(\s) \,\epsilon\,  x_-(\s)^{-1} \Bigr) - \frac{1}{2i} R^-\Bigl( x_+(\s) \,\epsilon\,  x_+(\s)^{-1} \Bigr).
\end{equation}
If we also introduce
\begin{equation*}
Z(\s) = \frac{1}{2i} \Bigl( x_+(\s) \,\epsilon\, x_+(\s)^{-1} -  x_-(\s) \,\epsilon\, x_-(\s)^{-1} \Bigr),
\end{equation*}
then one checks that
\begin{equation*}
\delta_\epsilon X(\s) = \bigl[ X(\s), K(\s) \bigr] = -\frac{1}{\gamma} \Bigl( \p_\s Z(\s) + \bigl[ X(\s), Z(\s) \bigr]_R \Bigr).
\end{equation*}
Using this identity and equation \eqref{dPExp}, we find that the transformation law of $x_\pm$ reads
\begin{equation*}
\delta_\epsilon x_\pm(\s) = R^\pm Z(\s) \, x_\pm(\s).
\end{equation*}
Finally, it follows that the pair of fields $\Psi_\pm$ simply transform as
\begin{equation*}
\delta_\epsilon \Psi_\pm(\s) = \Psi_\pm(\s) \epsilon.
\end{equation*}

It was observed in \cite{Delduc:2013fga} that for Yang-Baxter type deformations with standard $R$-matrices, the Cartan part of the $G$-symmetry is preserved. This can be checked here explicitly: indeed, for $\epsilon \in \h$ we find that the definition \eqref{DefK} of $K$ reduces to $\epsilon$, so that the infinitesimal transformation in the Cartan direction remains undeformed, as in \eqref{TransfogXNonDef}.
This fact can also be seen in terms of Poisson-Lie actions. For standard $R$-matrices, the Sklyanin bracket \eqref{SklyaninAlg} vanishes when restricted to the Cartan subalgebra $\h$. The corresponding action is then a usual Hamiltonian symmetry.

\subsubsection{Poisson-Lie symmetry: variation of the Hamiltonian and first order action}

In this section, we consider the case of the Yang-Baxter $\sigma$-model. The conservation of $T^g(\lambda_\pm)$ can be seen as the fact that it has a vanishing Poisson bracket with the Hamiltonian $H$ of the model. This implies that the Hamiltonian is invariant under the Poisson-Lie action generated by $T^g(\lambda_\pm)$, namely
\begin{equation*}
\delta_\epsilon H = 0.
\end{equation*}
Thus, the transformation \eqref{TransfogX} is a symmetry of the Hamiltonian.

Let us now compute the variation of the first order action under the transformation. In the case of a Hamiltonian action ($G^*$ abelian), the transformation is canonical and the invariance of the Hamiltonian is then equivalent to the invariance of the action. The situation is slightly more involved in the case of a Poisson-Lie action. The first order action is given by
\begin{equation}
S = \int \dd\tau\dd\s \, \kappa\bigl( g^{-1} \p_\tau g, X \bigr) - \int \dd\tau \, H.
\end{equation}
Consider the transformation \eqref{TransfogX} of $g$ and $X$ and, at first, let us allow the parameter $\epsilon$ to be a function of the time parameter $\tau$ with compact support. Since $H$ is invariant under this transformation, the variation of the action becomes
\begin{equation*}
\delta_\epsilon S = \int\dd\tau\dd\s \, \delta_\epsilon \Bigl( \kappa\bigl( g^{-1} \p_\tau g, X \bigr) \Bigr).
\end{equation*}
We have $\delta_\epsilon \bigl( g^{-1}\p_\tau g \bigr) = \p_\tau K + \bigl[ g^{-1}\p_\tau g, K \bigr]$, so that
\begin{equation*}
\delta_\epsilon \Bigl( \kappa\bigl( g^{-1} \p_\tau g, X \bigr) \Bigr) = \kappa \bigl( \p_\tau K, X \bigr) = \p_\tau \Bigl( \kappa(K,X) \Bigr) - \kappa\bigl( K, \p_\tau X \bigr).
\end{equation*}
Using expression \eqref{DefK2} for $K$, the skew-symmetry of $R$ and the invariance of $\kappa$ under the adjoint action, one finds
\begin{equation*}
\kappa\bigl( K, \p_\tau X \bigr) = \frac{1}{2i} \kappa \bigl( \epsilon, x_+^{-1} \p_\tau(R^+X) x_+ - x_-^{-1} \p_\tau(R^-X) x_- \bigr).
\end{equation*}
Discarding the boundary terms at initial and final times, we get
\begin{align*}
\delta_\epsilon S &= -\int\dd\tau \, \frac{1}{2i\gamma} \kappa \left( \epsilon, T^g(\lambda_+)^{-1} \int_{-\infty}^\infty \dd\sigma \, T^g(\lambda_+ \, ; +\infty, \s) \p_\tau \Lc^g(\lambda_+,\s) T^g(\lambda_+ \, ; \s, -\infty) \right) \\
&\qquad + \int\dd\tau \, \frac{1}{2i\gamma} \kappa \left( \epsilon, T^g(\lambda_-)^{-1} \int_{-\infty}^\infty \dd\sigma \, T^g(\lambda_- \, ; +\infty, \s) \p_\tau \Lc^g(\lambda_-,\s) T^g(\lambda_- \, ; \s, -\infty) \right).
\end{align*}
Using equation \eqref{dPExp}, this may be rewritten as
\begin{equation}
\delta_\epsilon S = -\frac{1}{2i\gamma} \int \dd\tau \, \kappa \bigl( \epsilon, T^g(\lambda_+)^{-1}\p_\tau T^g(\lambda_+)-T^g(\lambda_-)^{-1}\p_\tau T^g(\lambda_-) \bigr).
\end{equation}
In terms of the ``abstract'' non-abelian moment map $\Gamma$, seen as a $G^*$-valued map, this is simply
\begin{equation}
\delta_\epsilon S = -\int \dd\tau \,  \bigl\langle \epsilon(\tau), \Gamma^{-1}\p_\tau \Gamma \bigr\rangle.
\end{equation}
By the principle of least action, $\delta_\epsilon S$ must be zero for any function $\epsilon(\tau)$, as long as the fields are on-shell. Thus, we recover the fact that $\Gamma$ is conserved.

It is worth noticing that, if $G^*$ is non-abelian, $\Gamma^{-1} \p_\tau \Gamma$ is not a total time derivative. Thus, when we choose a constant parameter $\epsilon$, we cannot conclude that $\delta_\epsilon S=0$. That is to say, the action is not invariant under the Poisson-Lie symmetry. However, the latter is still a symmetry of the model since the Hamiltonian is invariant.

\section{Conclusion}

A deformation of Yang-Baxter type, with respect to either a split or non-split $R$-matrix, has the effect of $q$-deforming a global $G$-symmetry of the original integrable models.
In this article we showed at the Hamiltonian level that such a $q$-deformation can be understood as a deformation of the Poisson structure on the symmetry group $G$, from the trivial one to a multiple of the Sklyanin bracket defined by the $R$-matrix. Indeed, the non-abelian moment map $\Gamma : M \to G^\ast$ which generates this Poisson-Lie symmetry is given explicitly in terms of the monodromy matrix evaluated at the poles of the twist function, and its Poisson bracket is found to coincide with the pullback of the Semenov-Tian-Shansky bracket on $G^\ast \simeq G_R$. We then showed, in the case of a standard $R$-matrix, that the various charges constituting $\Gamma$ span the Poisson algebra $\U_q(\g)$.

\medskip

It is expected that Yang-Baxter type models possess a larger infinite dimensional symmetry which we could call $\U_q(\widehat \g)$, defined as a semiclassical limit of the quantum affine algebra $U_{\widehat q}(\widehat \g)$ with $\widehat q = q^\hbar$. This idea has indeed been realised explicitly in \cite{Kawaguchi:2012ve} where the Yang-Baxter $\sigma$-model for $\mathfrak{su}(2)$ was shown to have a $\U_q\big( \widehat{\mathfrak{su}(2)} \big)$ symmetry (see also \cite{Kawaguchi:2013gma}). However, even in these rank one cases, deriving the full infinite dimensional symmetry algebra requires using the algebra of the monodromy matrix for arbitrary values of the spectral parameters, as opposed to its evaluation at the poles of the twist function. Yet the former is notoriously plagued with ambiguities due to the non-ultralocality in the Poisson bracket of the Lax matrix of these models.

\medskip

Another important class of integrable deformations of integrable $\sigma$-models, 
which applies in particular to the principal chiral model and (semi)-symmetric 
space $\sigma$-models, is given by the $\lambda$-deformations (sometimes also 
called $k$-deformations or deformations of the gauged WZW type) 
\cite{Sfetsos:2013wia,Hollowood:2014rla,Hollowood:2014qma,Sfetsos:2015nya},    
see also \cite{Itsios:2014vfa, 
Sfetsos:2014lla,Demulder:2015lva,Vicedo:2015pna,Hoare:2015gda,Klimcik:2015gba,
Hollowood:2015dpa,Appadu:2015nfa,Thompson:2015lzd}.
  There are results \cite{Evans:1994hi,Hollowood:2015dpa} indicating 
 that these deformed models possess a $\U_q(\g)$ symmetry with $q$ a phase. It would therefore be interesting to also relate this 
type of $q$-deformation to Poisson-Lie symmetries. However, the study of the symmetries 
of these models along the lines of the present article is more difficult. For instance, 
contrary to the setting of section \ref{sec: YB models} for Yang-Baxter type models 
where the Poisson bracket of the quantity $T^g(\lambda_\pm)$ with itself is well defined, it 
appears that one would have to deal with the issue of non-ultralocality.

\paragraph{Acknowedgements.} 
We thank K. Gawedzki for useful discussions and T. J. Hollowood for comments on the draft. 
This work is partially supported by the program PICS 6412 DIGEST of CNRS and by the French Agence Nationale de la Recherche (ANR) under grant ANR-15-CE31-0006 DefIS.

\appendix

\section{Semisimple complex Lie algebras}
\label{App:SemiSimpleStruct}

Let $\f$ be a semisimple complex Lie algebra. Fix a Cartan subalgebra $\h$ of $\f$ and let $\Delta \subset \h^*$ denote the associated set of roots. We choose a set of simple roots $\alpha_1,\ldots,\alpha_l \in \Delta$, with $l$ the rank of $\f$. Let $\f=\h\oplus\n_+\oplus\n_-$ be the associated Cartan-Weyl decomposition, and $\lbrace E_\alpha, \alpha >0 \rbrace$ and $\lbrace E_{-\alpha}, \alpha >0 \rbrace$ the associated bases of the nilpotent subalgebras $\n_\pm$.
We choose to normalise the latter such that
\begin{equation*}
\kappa\left(E_\alpha,E_\beta\right)=\delta_{\alpha,-\beta},
\end{equation*}
where $\kappa$ is the Killing form on $\f$. For $\alpha\in\Delta$, we define $H_\alpha \in \h$ \textit{via} the Killing form isomorphism between $\h$ and $\h^*$, namely we set
\begin{equation} \label{Halpha def}
\kappa(H_\alpha,X) = \alpha(X).
\end{equation}
for all $X\in\h$. In particular, for simple roots $\alpha_i$ we use the shorthand notation $H_i=H_{\alpha_i}$. Then, $\lbrace H_i, \, i=1,\ldots,l \rbrace$ forms a basis of $\h$. The Killing form induces a symmetric bilinear form on $\h^*$ that we shall note $(\cdot,\cdot)$.

By definition, we have the following commutation relations
\begin{equation}
\left[X, E_\alpha \right] = \alpha(X) E_\alpha. 
\end{equation}
for all $X\in\h$. Moreover,
\begin{equation}\label{ComEF}
\left[E_{\alpha}, E_{-\alpha}\right]=H_\alpha \; \; \; \; \text{and} \; \; \; \; \left[E_{\alpha_i},E_{-\alpha_j}\right] = \delta_{ij}H_i.
\end{equation}
Finally, the Lie algebra structure of $\n_\pm$ is given by
\begin{equation}\label{ComEE}
\left[ E_\alpha, E_\beta \right] = N_{\alpha,\beta} E_{\alpha+\beta},
\end{equation}
with $N_{\alpha,\beta}$ a real skew-symmetric normalisation constant. Moreover, one has $N_{-\alpha,-\beta}=-N_{\alpha,\beta}$.

\section{Split and non-split real forms}
\label{App:RealForms}

Let $\f$ be a semisimple complex Lie algebra of dimension $n$. We will use the notations introduced in appendix \ref{App:SemiSimpleStruct}. We want to describe the real forms of $\f$, \textit{i.e.} the subalgebras of $\f$ which are also real vector space of dimension $n$.

The real forms of $\f$ are in one-to-one correspondence with the fixed-point subalgebras $\f^\theta$ of semi-linear involutive automorphisms $\theta$ of $\f$, \textit{i.e.} maps $\theta:\f\rightarrow\f$ such that $\theta^2=\Id$ and
\begin{equation*}
\theta(\lambda X + \mu Y) = \overline{\lambda} \theta(X) + \overline{\mu}\theta(Y) \; \; \text{and} \; \; \theta\bigl([X,Y]\bigr)=\bigl[\theta(X),\theta(Y)\bigr],
\end{equation*}
for all $\lambda, \mu \in \C$ and $X,Y \in \f$.

Consider the Cartan-Weyl basis $\lbrace H_i, E_\alpha \rbrace$ of $\f$ (cf. appendix \ref{App:SemiSimpleStruct}). Since the $E_{\pm\alpha_i}$'s generate the Lie algebra $\f$, the automorphism $\theta$ is completely described by its action on these. In the following, we consider two ``natural'' ways to act on the $E_{\pm\alpha_i}$'s, that correspond to the so-called split and non-split real forms.

\paragraph{Split real form.}
Let us consider the semi-linear involutive automorphism $\theta$ defined by
\begin{equation}
\theta \left(E_{\pm\alpha_i} \right) = E_{\pm\alpha_i}.
\end{equation}
As the normalisation constants $N_{\alpha,\beta}$ in equation \eqref{ComEE} are all real, we find for any root $\alpha$ that
\begin{equation*}
\theta \left( E_\alpha \right) = E_{\alpha}.
\end{equation*}
Hence, one also has
\begin{equation*}
\theta \left( H_\alpha \right) = H_\alpha.
\end{equation*}
As a consequence, a basis of the real subalgebra fixed by $\theta$ is given by the Cartan-Weyl basis $\lbrace H_i, E_\alpha \rbrace$ of $\f$ itself. This way, we obtain the so-called split real form
\begin{equation}\label{BasisSplit}
\g = \left\lbrace \sum_{i=1}^l a_i H_i + \sum_{\alpha\in\Delta} b_\alpha E_\alpha, \; a_i, b_\alpha \in \mathbb{R} \right\rbrace.
\end{equation}

\paragraph{Non-split real forms.}
Another possibility for defining $\theta$ is to let
\begin{equation}
\theta \left(E_{\pm\alpha_i} \right) = -\lambda_i E_{\mp\alpha_i},
\end{equation}
where $\lambda_i=\pm 1$. Using equation \eqref{ComEE}, for any positive root $\alpha=p_1\alpha_1+\ldots+p_l\alpha_l$ we obtain
\begin{equation}
\theta \left(E_\alpha \right) = -\lambda_{\alpha} E_{-\alpha},
\end{equation}
with $\lambda_\alpha= \lambda_1^{p_1}\ldots\lambda_l^{p_l} \, \in \lbrace +1,-1 \rbrace$. In the same way, using \eqref{ComEF}, one has
\begin{equation}
\theta(H_\alpha)=- H_\alpha.
\end{equation}
The real subalgebra $\g$ of elements fixed by $\theta$ is called a non-split real form of $\f$. A basis of $\g$ is given by
\begin{equation}\label{BasisNonSplit}
T_i=iH_i, \; \; \; B_\alpha = \frac{i}{\sqrt{2}} \left(E_\alpha+\lambda_\alpha E_{-\alpha} \right), \; \; \; C_\alpha = \frac{1}{\sqrt{2}} \left(E_\alpha-\lambda_\alpha E_{-\alpha} \right).
\end{equation}
If one choose $\lambda_1=\ldots=\lambda_l=1$, then $\lambda_\alpha=1$ for any root $\alpha$ and we get the so-called compact real form of $\f$.

\section{Roots, co-roots, weights and co-weights}
\label{App:BasesCartan}

In this appendix, we recall the main properties of some bases of the Cartan subalgebra $\h$. As it is equipped with the non-degenerate Killing form $\kappa(\cdot,\cdot)$, there exists a natural isomorphism
\begin{equation*}
\zeta : \h^* \rightarrow \h,
\end{equation*}
between $\h$ and its dual $\h^*$. It is characterised by the relation
\begin{equation*}
\kappa\bigl( \zeta(\lambda),X \bigr) = \lambda(X),
\end{equation*}
for all $\lambda \in \h^*$ and $X \in \h$ (cf. \eqref{Halpha def}).
This isomorphism induces a bilinear form $(\cdot,\cdot)$ on $\h^*$, given for all $\lambda,\mu \in \h^*$ by
\begin{equation*}
( \lambda, \mu ) = \kappa\bigl( \zeta(\lambda), \zeta(\mu) \bigr).
\end{equation*}

\subsection{Roots and co-roots}

A basis of $\h^*$ is given by the simple roots $\alpha_1,\ldots,\alpha_l$. Using the notations of appendix \ref{App:SemiSimpleStruct}, the corresponding basis $\lbrace \zeta(\alpha_i) \rbrace$ of $\h$ is $\lbrace H_1,\ldots,H_l \rbrace$. We will often use
\begin{equation*}
d_i = \frac{(\alpha_i,\alpha_i)}{2} = \frac{\kappa(H_i,H_i)}{2}.
\end{equation*}
We define the simple co-roots as 
\begin{equation*}
\alpha_i^\vee = \frac{2H_i}{(\alpha_i,\alpha_i)}=d_i^{-1} H_i,
\end{equation*}
which forms another basis of $\h$. The Cartan matrix is then given by
\begin{equation*}
A_{ij}=\alpha_j(\alpha_i^\vee)=\frac{2(\alpha_i,\alpha_j)}{(\alpha_i,\alpha_i)}.
\end{equation*}

\subsection{Fundamental weights and co-weights}

We define the fundamental weights $\omega_i \in \h^*$ as the dual basis of the co-roots $\alpha_i^\vee$, namely
\begin{equation*}
\omega_i(\alpha_j^\vee)=\delta_{ij}.
\end{equation*}
By the Killing form duality, the
\begin{equation} \label{def Pi}
P_i=\zeta(\omega_i)
\end{equation}
form a basis of $\h$. Moreover, we have the relation
\begin{equation}\label{AlphaP}
\alpha_i(P_j) = d_i \delta_{ij}.
\end{equation} 
In the same way, one defines the fundamental co-weights $\omega_i^\vee \in \h$ as the dual basis of the simple roots
\begin{equation*}
\alpha_j(\omega_i^\vee)=\delta_{ij},
\end{equation*}
which relates them to the basis elements \eqref{def Pi} simply by
\begin{equation*}
\omega_i^\vee=d_i^{-1}P_i.
\end{equation*}

\section{Poisson brackets extraction theorems}
\label{App:ThmPB}

In this appendix, we state and prove two theorems allowing to extract the Poisson brackets of the factors of a Lie-group-valued quantity.

\begin{theorem}\label{TheoremPB1}
Let $F_1$ and $F_2$ be two Lie groups, that are decomposable into two subgroups: $F_i=G_iH_i$. This group factorisation corresponds to a direct sum of Lie algebras $\f_i = \g_i \oplus \h_i$. We denote by $\pi_{\g_i}$ and $\pi_{\h_i}$ the associated projections. We consider $A \in F_1$ and $B \in F_2$ that we factorise as
\begin{equation*}
A=uv, \; \; \; \text{and} \: \: \: B=xy,
\end{equation*}
for $(u,v) \in G_1 \times H_1$ and $(x,y) \in G_2 \times H_2$. We define
\begin{equation*}
\Pc\ti{12} = u^{-1}\ti{1}x^{-1}\ti{2} \lbrace A\ti{1}, B\ti{2} \rbrace v^{-1}\ti{1}y^{-1}\ti{2} \; \in \f_1 \otimes \f_2.
\end{equation*}
Then, we have
\begin{align*}
\lbrace u\ti{1}, x\ti{2} \rbrace &= u\ti{1}x\ti{2} \; \pi_{\g_1} \otimes \pi_{\g_2} \bigl( \Pc\ti{12} \bigr), \\
\lbrace u\ti{1}, y\ti{2} \rbrace &= u\ti{1} \; \pi_{\g_1} \otimes \pi_{\h_2} \bigl( \Pc\ti{12} \bigr) \; y\ti{2}, \\
\lbrace v\ti{1}, x\ti{2} \rbrace &= x\ti{2} \; \pi_{\h_1} \otimes \pi_{\g_2} \bigl( \Pc\ti{12} \bigr) \; v\ti{1}, \\\lbrace v\ti{1}, y\ti{2} \rbrace &= \pi_{\h_1} \otimes \pi_{\h_2} \bigl( \Pc\ti{12} \bigr) \; v\ti{1}y\ti{2}.
\end{align*}
\begin{proof}
Using the Leibniz rule, we have
\begin{equation*}
\lbrace A\ti{1}, B\ti{2} \rbrace = \lbrace u\ti{1}, x\ti{2} \rbrace v\ti{1}y\ti{2} + x\ti{2}\lbrace u\ti{1}, y\ti{2} \rbrace v\ti{1} + u\ti{1}\lbrace v\ti{1}, x\ti{2} \rbrace y\ti{2} + u\ti{1}x\ti{2}\lbrace v\ti{1}, y\ti{2} \rbrace,
\end{equation*}
so that
\begin{equation*}
\Pc\ti{12} = 
\underbrace{u^{-1}\ti{1}x^{-1}\ti{2} \lbrace u\ti{1}, x\ti{2} \rbrace}_{\displaystyle \in \g_1 \otimes \g_2} + 
\underbrace{u^{-1}\ti{1} \lbrace u\ti{1}, y\ti{2} \rbrace y^{-1}\ti{2}}_{\displaystyle \in \g_1 \otimes \h_2} +
\underbrace{x^{-1}\ti{2} \lbrace v\ti{1}, x\ti{2} \rbrace v^{-1}\ti{1}}_{\displaystyle \in \h_1 \otimes \g_2} +
\underbrace{\lbrace v\ti{1}, y\ti{2} \rbrace v^{-1}\ti{1} y^{-1}\ti{2}}_{\displaystyle \in \h_1 \otimes \h_2},
\end{equation*}
and hence the theorem.
\end{proof}
\end{theorem}

\begin{theorem}\label{TheoremPB2}
Let $G$ be a Lie group that factorises into subgroups as $G=G_1 \ldots G_p$. This group factorisation corresponds to a direct sum of Lie algebras $\g = \g_1 \oplus \ldots \oplus \g_p$, with associated projections $\pi_i$. Suppose this decomposition is such that, for all $i\in\lbrace 1,\ldots,p \rbrace$,
\begin{equation*}
\g_{<i}=\bigoplus_{k=1}^{i-1} \g_k \; \; \; \; \text{ and } \; \; \; \; \g_{>i}=\bigoplus_{k=i+1}^p \g_k
\end{equation*}
are subalgebras of $\g$ and $[\g_i,\g_{<i}] \subseteq \g_{<i}$.

Let $f$ be a $\mathbb{R}$-valued function and $A$ a $G$-valued function, that we factorise as
\begin{equation*}
A=A^{(1)} \ldots A^{(p)}, \; \; \; \text{with } \left(A^{(1)},\ldots,A^{(p)}\right) \in G_1 \times \ldots \times G_p.
\end{equation*}
If we define
\begin{equation*}
\Pc^{(i)} = \bigl( A^{(1)} \ldots A^{(i)} \bigr)^{-1} \left\lbrace A, f \right\rbrace \bigl( A^{(i+1)} \ldots A^{(p)} \bigr)^{-1} \in \g,
\end{equation*}
then we have
\begin{equation*}
\left(A^{(i)}\right)^{-1}\left\lbrace A^{(i)}, f \right\rbrace = \pi_i  \bigl( \Pc^{(i)} \bigr).
\end{equation*}
\begin{proof}
Let $B = A^{(1)} \ldots A^{(i-1)}$ and $C = A^{(i+1)} \ldots A^{(p)}$. Using the Leibniz rule, we have
\begin{equation*}
\Pc^{(i)} = \underbrace{\left(A^{(i)}\right)^{-1} \left\lbrace A^{(i)}, f \right\rbrace}_{\in \g_i} + \left(A^{(i)}\right)^{-1} \underbrace{\bigl( B^{-1} \left\lbrace B, f \right\rbrace \bigr) }_{\in \g_{<i}}A^{(i)} + \underbrace{\lbrace C, f \rbrace C^{-1}}_{\in\g_{>i}}.
\end{equation*}
Since $B^{-1} \left\lbrace B, f \right\rbrace$ belongs to $\g_{<i}$, the assumption on the Lie subalgebras $\g_k$ tells us that the adjoint action
\begin{equation*}
\left(A^{(i)}\right)^{-1}\bigl( B^{-1} \left\lbrace B, f \right\rbrace \bigr)A^{(i)}
\end{equation*}
still belongs to $\g_{<i}$, hence the theorem.
\end{proof}
\end{theorem}

\section{Path-ordered exponentials}
\label{App:PathExp}

In this appendix, we recall some properties of path-ordered exponentials. Consider a $\g$-valued field $\Lc(\s)$ and the path-ordered exponential
\begin{equation} \label{T Pexp def}
T(\s,\s')= \Pexp \left( -\int_{\s'}^\s \dd\rho \, \Lc(\rho) \right).
\end{equation}
This is a $G$-valued field verifying the differential equations
\begin{align*}
\bigl( \p_\s T(\s,\s') \bigr) T(\s,\s')^{-1} &= -\Lc(\s), \\
T(\s,\s')^{-1} \bigl( \p_{\s'} T(\s,\s') \bigr) &= \Lc(\s').
\end{align*}
Under an infinitesimal transformation $\delta \Lc$ of $\Lc$, the path-ordered exponential is transformed by
\begin{equation}\label{dPExp}
\delta T(\s,\s') = - \int_{\s'}^\s \dd\rho \; T(\s,\rho) \delta\Lc(\rho) T(\rho,\s').
\end{equation}
This formula allows one to compute the variations of $T(\s,\s')$ or Poisson brackets of $T(\s,\s')$ with other observables.

In particular, if $\Lc$ is the spatial component of a zero curvature equation
\begin{equation*}
\p_\tau \Lc - \p_\s \mathcal{M} + [\mathcal{M},\mathcal{L}] = 0,
\end{equation*}
we find from equation \eqref{dPExp} that
\begin{equation*}
\p_\tau T(\s,\s') = T(\s,\s')\mathcal{M}(\s')-\mathcal{M}(\s)T(\s,\s').
\end{equation*}
Thus, if we consider fields that decrease rapidly at infinity, so that $\mathcal{M}(\s) \rightarrow 0$ when $\s \rightarrow \pm\infty$, the path-ordered exponential $T(+\infty,-\infty)$ is conserved.

Suppose now that we are given a Lie group homomorphism $\tau: G \rightarrow F$, from $G$ to another Lie group $F$. It induces a Lie algebra homomorphism $\tau_\g : \g \rightarrow \f$. The image of the path-ordered exponential \eqref{T Pexp def} by the homomorphism $\tau$ is simply
\begin{equation}\label{AutoPExp}
\tau \bigl( T(\s,\s') \bigr) = \Pexp_F \left( -\int_{\s'}^\s \dd\rho \, \tau_\g \bigl( \Lc(\rho) \bigr) \right),
\end{equation}
where $\Pexp_F$ denotes the path-ordered exponential in the group $F$.

 \providecommand{\href}[2]{#2}\begingroup\raggedright\endgroup

\end{document}